\documentclass{emulateapj}

\usepackage{amsmath}
\usepackage{amssymb}
\usepackage{graphicx}
\usepackage{color}
\usepackage{natbib}
\usepackage{epsfig}

\def\gtaprx {\lower .1ex\hbox{\rlap{\raise .6ex\hbox{\hskip .3ex
	{\ifmmode{\scriptscriptstyle >}\else
		{$\scriptscriptstyle >$}\fi}}}
	\kern -.4ex{\ifmmode{\scriptscriptstyle \sim}\else
		{$\scriptscriptstyle\sim$}\fi}}}
\def\ltaprx {\lower .1ex\hbox{\rlap{\raise .6ex\hbox{\hskip .3ex
	{\ifmmode{\scriptscriptstyle <}\else
		{$\scriptscriptstyle <$}\fi}}}
	\kern -.4ex{\ifmmode{\scriptscriptstyle \sim}\else
		{$\scriptscriptstyle\sim$}\fi}}}

\newcommand{\cutt}[1]{\textcolor{blue}{}}

\newcommand{\Ms}{{\ensuremath{{M}_{\odot} }}}
\newcommand{\Zs}{\ensuremath{Z_\odot}}

\newcommand{\Ni}{{\ensuremath{^{56}\mathrm{Ni}}}}

\begin{document}

\title{Finding the First Cosmic Explosions II: Core-Collapse Supernovae}

\author{Daniel J. Whalen\altaffilmark{1,2},  Candace C. Joggerst\altaffilmark{2}, Chris 
L. Fryer\altaffilmark{3}, Massimo Stiavelli\altaffilmark{4}, Alexander Heger\altaffilmark{5}
and Daniel E. Holz\altaffilmark{6}}

\altaffiltext{1}{McWilliams Fellow, Department of Physics, Carnegie Mellon 
University, Pittsburgh, PA 15213}

\altaffiltext{2}{T-2, Los Alamos National Laboratory, Los Alamos, NM 87545}

\altaffiltext{3}{CCS-2, Los Alamos National Laboratory, Los Alamos, NM 87545}

\altaffiltext{4}{Space Telescope Science Institute, 3700 San Martin Drive, Baltimore,
MD 21218}

\altaffiltext{5}{School of Physics and Astronomy, University of Minnesota,
Minneapolis, MN  55455}

\altaffiltext{6}{Enrico Fermi Institute, Department of Physics, and Kavli Institute for 
Cosmological Physics, University of Chicago, Chicago, IL 60637, USA}

\begin{abstract}

Understanding the properties of Pop III stars is prerequisite to elucidating the nature of 
primeval galaxies, the chemical enrichment and reionization of the early IGM, and the 
origin of supermassive black holes.  While the primordial IMF remains unknown, recent 
evidence from numerical simulations and stellar archaeology suggests that some Pop III 
stars may have had lower masses than previously thought, 15 - 50 \Ms\ in addition to 50 
- 500 \Ms.  The detection of Pop III supernovae by \textit{JWST}, \textit{WFIRST} or the 
\textit{TMT} could directly probe the primordial IMF for the first time.  We present numerical 
simulations of 15 - 40 \Ms\ Pop III core-collapse SNe done with the Los Alamos radiation 
hydrodynamics code RAGE.  We find that they will be visible in the earliest galaxies out 
to $z \sim$ 10 - 15, tracing their star formation rates and in some cases revealing their 
positions on the sky. Since the central engines of Pop III and solar-metallicity core-collapse 
SNe are quite similar, future detection of any Type II supernovae by next-generation NIR 
instruments will in general be limited to this epoch.

\vspace{0.1in}

\end{abstract}


\section{Introduction}

The first stars are crucial to the formation of primeval galaxies \citep{jgb08,get08,jlj09,
get10,jeon11,pmb11,pmb12,wise12}, the chemical enrichment of the early IGM \citep{
mbh03,ss07,bsmith09,chiaki12,ritt12}, the initial stages of cosmological reionization 
\citep{wan04,ket04,abs06,awb07,wa08a}, and the origin of supermassive black holes 
\citep{bl03,jb07b,brmvol08,milos09,awa09,lfh09,th09,li11,pm11,pm12,jlj12a,wf12,
agarw12,jet13,pm13}.  Unfortunately, because they lie beyond the reach of current 
ground and space based instruments there are not yet any observational constraints 
on their properties.  The early numerical simulations of Pop III stars suggest that they 
form in isolation, one per halo, in 10$^5$ - 10$^6$ \Ms\ dark matter halos at  $z \sim$ 
20 - 30 and are very massive, 100 - 500 \Ms\ \citep{bcl99,abn00,abn02, bcl02,nu01,
y08}.  Such stars usually drive most of the baryons from their halos in strong ionized 
flows that create diffuse H II regions with $n \sim$ 0.1 - 1 cm$^{-3}$ \citep{wan04,ket04,
abs06,awb07,wa08a,wn08a,wn08b}. Pop III stars are not thought to lose much mass 
over their lifetimes because there are no line-driven winds in their metal-free 
atmospheres \citep{Kudritzki00,Vink01,Baraffe01,kk06,Ekstr08}.

There is growing evidence that some Pop III stars may be less massive than previously 
thought.  Recent, more extensive ensembles of numerical simulations have found many 
halos with central collapse rates consistent with 20 - 60 \Ms\ for the final mass of the star 
\citep{on07,wa07,on08} and that a fraction of the halos form binaries in this mass range 
\citep{turk09}.  New simulations of Pop III protostellar accretion disks suggest that they 
were prone to fragmentation into as many as a dozen smaller stars \citep{stacy10,clark11,
sm11,get11,get12}. Preliminary models also suggest that ionizing UV breakout from 
primordial protostellar disks limits the masses of some Pop III stars to $\sim$ 40 \Ms\ 
\citep{hos11,stacy12,hos12} \citep[although also see][]{op01,oi02,op03,tm04,tm08}.  

While these new models are important steps forward they cannot predict the final masses 
of primordial stars. For example, the fragmentation of Pop III protostellar disks has only 
been followed out to a few centuries, far short of the time required to build up a massive 
star, and many of the fragments in these models are later found to merge with the central 
object instead of becoming stars themselves.  Thus, it is not clear if a small swarm of less 
massive stars forms or a single, very massive star is created, albeit through protracted 
clumpy accretion.  Clumpy accretion may also keep the protostar puffed up and cool, 
allowing it to reach much higher masses before evaporating its accretion disk than in the 
\citet{hos11} models.  Because no simulation realistically bridges the gap between the 
formation and fragmentation of the accretion disk and its destruction up to a Myr later, 
they cannot yet constrain the Pop III initial mass function \citep[IMF; for recent reviews,
see][]{glov12,dw12}.  We also note that the effects of turbulence \citep{schl13}, magnetic 
fields due to the small scale turbulent dynamo \citep{schob12}, and radiation transport on 
the evolution of the disk are not well understood.
 
There have been attempts to determine the masses of the first stars by comparing the 
cumulative nucleosynthetic yield of their supernovae (SNe) to the fossil abundance record, 
the chemical abundances measured in ancient, dim extremely metal-poor (EMP) stars out 
in the Galactic halo \citep[e.g.][]{bc05,fet05,caffau12}.  Pop III stars from 15 - 40 \Ms\ die in 
core-collapse (CC) SN explosions and 140 - 260 \Ms\ stars explode as far more energetic 
pair-instability (PI) SNe \citep{hw02} \citep[although the lower mass limit for PI SNe has 
recently dropped down to 65 \Ms\ for rotating progenitors --][]{cw12}.  \citet{jet09b} 
(hereafter JET10) found that the chemical yields of 15 - 40 \Ms\ Pop III SNe are a good 
match to the abundances observed in a sample of $\sim$ 130 EMP stars \citep{Cayrel2004,
Lai2008}, and that such stars may have been responsible for much of the chemical 
enrichment of the primeval IGM.  However, evidence for the "odd-even" nucleosynthetic 
imprint of Pop III pair-instability supernovae (PI SNe) has now been found in high redshift 
damped Lyman alpha absorbers \citep{cooke11}, and there are indications it could be
pesent in a new sample of 18 metal-poor stars in the \textit{Sloan Digital Sky Survey} \citep{
ren12} \citep[see also][on why the odd-even effect may not have been discovered in earlier 
surveys]{karl08}.  Together, these discoveries suggest that both high-mass and low-mass 
Pop III star formation was possible.  In sum, while stellar archaeology has revealed some 
insights into the first stellar populations it has not yet placed firm constraints on their masses.

Observations of primordial SNe \citep{byh03,vas12} could directly probe the Pop III IMF for 
the first time because in principle they can be detected at great distances and distinguish 
between low mass and high mass progenitors.  Until now, calculations of Pop III SN light 
curves and spectra have been confined to PI SNe to determine if they can be detected by 
future SN 1a missions \citep[$4 < z < 6$,][]{sc05} or by the \textit{James Webb Space 
Telescope} \citep[\textit{JWST},][]{jwst06} during the era of reionization \citep[$6<z<15$,
][]{kasen11,pan12a} or at $z \sim$ 30 \citep{wa05,hum12} \citep[see also][]{pan12b,det12}.  
Simulations of Pop III PI SN light curves have also been applied to PI candidates in the local 
universe such as SN 2007bi \citep{gy09,yn10}.  Most recently, radiation hydrodynamical 
models of Pop III PI SN light curves and spectra by \citet{wet12a,wet12b,wet12d} showed 
for the first time that these explosions will be visible out to $z \sim 30$ with \textit{JWST} and 
$z \sim$ 15 - 20 in all-sky near infrared (NIR) surveys such as the \textit{Wide-Field Infrared 
Survey Telescope} (\textit{WFIRST}) or the \textit{Wide-Field Imaging Surveyor for High 
Redshift} (\textit{WISH}). However, no calculation to date has addressed detection thresholds 
in redshift for Pop III CC SNe, whose numbers may be greater than those of PI SNe at early 
epochs.

We present radiation hydrodynamical simulations of light curves and spectra for 15 - 40 
\Ms\ Pop III explosions done with the Los Alamos RAGE and SPECTRUM codes.  In 
Section 2 we describe our grid of explosion models and how we post process them obtain 
light curves and spectra. In Section 3 we examine blast profiles and spectra for Pop III CC 
SNe in detail, and in Section 4 we compute their near infrared (NIR) light curves and 
detection thresholds in redshift.  In Section 5 we conclude.

\section{Numerical Algorithm}

We adopt the models in JET10 for our grid of light curve simulations because they 
span the range of progenitor masses, structures and explosion energies expected for 
such stars and because their elemental yields are a good match to those in the fossil 
abundance record.  We calculate light curves and spectra in four stages.  First, 15 - 
40 \Ms\ Pop III stars are evolved through all stages of stable nuclear burning and then 
exploded in the Kepler code.  Each explosion is then mapped onto a 2D adaptive mesh 
refinement (AMR) grid in the CASTRO code and evolved until just before the shock 
reaches the surface of the star to capture internal mixing prior to breakout.  We then 
spherically average our CASTRO profiles onto 1D AMR grids in RAGE and evolve 
them out to 4 months, after which the ejecta dims below observability.  Finally, we post 
process our RAGE profiles with the SPECTRUM code to construct light curves and 
spectra with detailed atomic opacity sets.  

\subsection{Kepler}

We use the 1D Lagrangian stellar evolution code Kepler \citep{Weaver1978,Woosley2002} 
to evolve the progenitors from the beginning of the main sequence to the onset of collapse 
of their iron cores.  The explosions are then artificially triggered by a piston at a constant 
Lagrangian mass coordinate that advances though the star with a specified radial history.  
The SNe are evolved through all nuclear burning ($\sim$100 s after the explosion) and 
then the simulations are halted, well before the shock exits the He shell and any reverse 
shocks form in which instabilities might develop.  We follow energy production with a 
19-isotope network until oxygen is depleted from the core and with a 128-isotope 
quasi-equilibrium network thereafter.  Stellar rotation is approximated by a semi-convective 
mixing parameter.  

As in JET10, we consider two progenitor metallicities ($Z=$ 0 and 10$^{-4}$ \Zs\ denoted 
by z and u, respectively), three explosion energies ($E_{ex}=$ 0.6, 1.2, and 2.4 B 
designated by B, D, and G, respectively, where 1 B = 1 Bethe = 10$^{51}$ erg), and three 
progenitor masses (15, 25, and 40 \Ms), a total of 18 models.  Thus, model z15G is the 
2.4 B explosion of a zero-metallicity 15 \Ms\ Pop III star.  We use only one rotation rate 
from that study (R = 5, where R is defined in $\S \, 3$ of JET10) because it was found that 
the rate of rotation had little effect on the degree of mixing during the explosion. In contrast 
to the u- and z-series 150 - 250 \Ms\ stars in \citet{wet12b}, u-series 15 - 40 \Ms\ Pop III 
stars die as compact blue giants and z-series stars die as red supergiants.

Further details on our Kepler models, such as our numerical mesh, explosion triggers, and 
use of semi-convective mixing as a proxy for rotation, can be found in Section 2 of JET10.  
Besides altering the envelope of the the star, rotation can also trigger asymmetries in the 
central explosion engine whose effects on the SN light curves are not explored here.  For 
example, jetlike instabilities could arise in the shock that lead to breakout times that vary 
with angle and alter the luminosity of the explosion.  In extreme cases, such jets can burn 
material in the core and form large amounts of \Ni\ that also cause the SN to be brighter at 
intermediate times \citep[e.g., hypernovae;][]{Iwamoto2005}.  Asymmetries in the explosion 
can be seeded prior to core bounce and cause even stronger instabilities and central mixing 
at later times.  The effects of these processes will be examined in future models.

\subsection{CASTRO}

As in JET10, we map our Kepler explosion profiles onto a 2D axisymmetric grid in the 
new CASTRO code \citep{Almgren2010} and evolve the shock up to the edge of the 
star in mass coordinate.  CASTRO is a multidimensional Eulerian AMR code with a high
order, unsplit Godunov hydrodynamics scheme.  We follow the same 27 elements from 
hydrogen to zinc as in JET10 and include the gravity of the compact remnant, which 
JET10 and others have shown to be crucial to both fallback and mixing in Pop III CC 
explosions \citep{zwh08}.  Radiation transport is not necessary in this stage of our 
calculations because photon mean free paths in the star prior to breakout are so short 
that they are simply advected through the star by the fluid flow.  We do account for the 
contribution of photons to pressure in the equation of state (EOS).  Our models also 
include energy deposition due to radioactive decay of \Ni\ in the ejecta, as described by 
equation 4 in JET10.

The JET10 models show that mixing is mostly complete by the time the shock reaches 
the surface of the star, so elemental mass fractions are realistically distributed in radius 
and angle in our profiles when we spherically average them and map them into RAGE.  
We halt the CASTRO simulation when the shock is no less than 100 photon mean free 
paths $\lambda_p$ from the edge of the star: \vspace{0.05in}
\begin{equation}
\lambda_p = \frac{1}{\kappa\rho}, \vspace{0.05in}
\end{equation}
where $\kappa$ is the opacity due to Thomson scattering from electrons, taken to be 
0.288 gm cm$^2$, and $\rho$ is the density just ahead of the shock inside the star.  

This intermediate step in CASTRO allows us to capture how heavy elements that have 
built up in the star are mixed in the initial stages of the SN without resorting to full 2D 
radiation transport models in RAGE.  Such calculations, while tractable, would require
too much time for a grid of models as large as ours and are not necessary because 
photon transport does not affect mixing. Mixing determines the order in which emission 
and absorption lines appear in the spectra over time.  If metal lines appear soon after 
shock breakout in a Pop III SN they would be a clear signature of a core-collapse event 
because only in such explosions can mixing dredge these elements up from deep inside 
the star.  Intermediate modeling in 2D is not needed for PI SN light curves because 
mixing is not vigorous enough to bring elements from the core up to the photosphere of 
the explosion at early times \citep{jw11,chen11}.

We refer our readers to $\S \, 2$ of JET10 for further details on our CASTRO numerical 
mesh, boundary conditions, gravity scheme, EOS, AMR refinement criteria, and energy 
deposition due to \Ni \ decay.

\subsection{RAGE}

\begin{figure*}
\begin{center}
\begin{tabular}{cc}
\epsfig{file=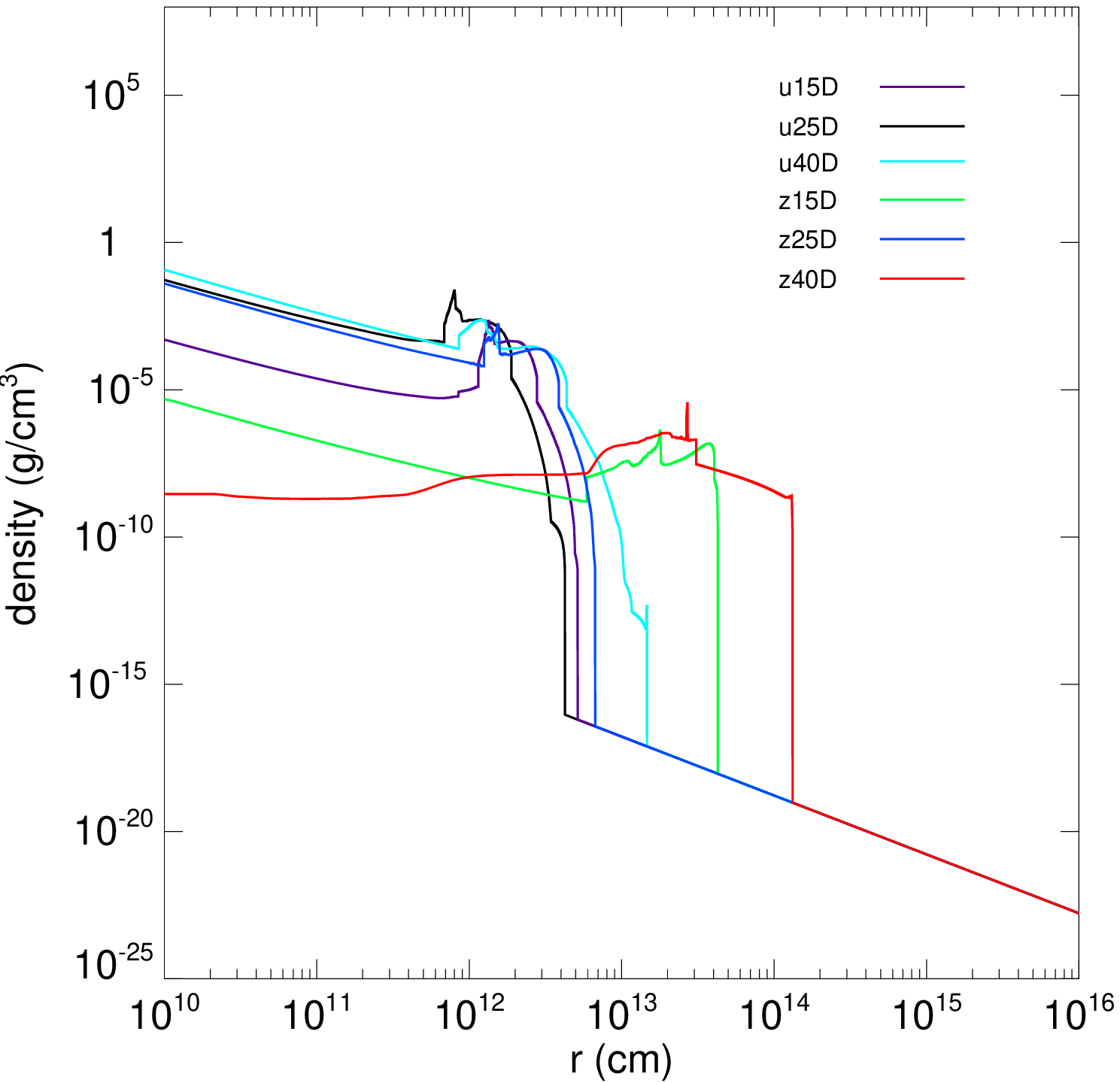,width=0.45\linewidth,clip=} & 
\epsfig{file=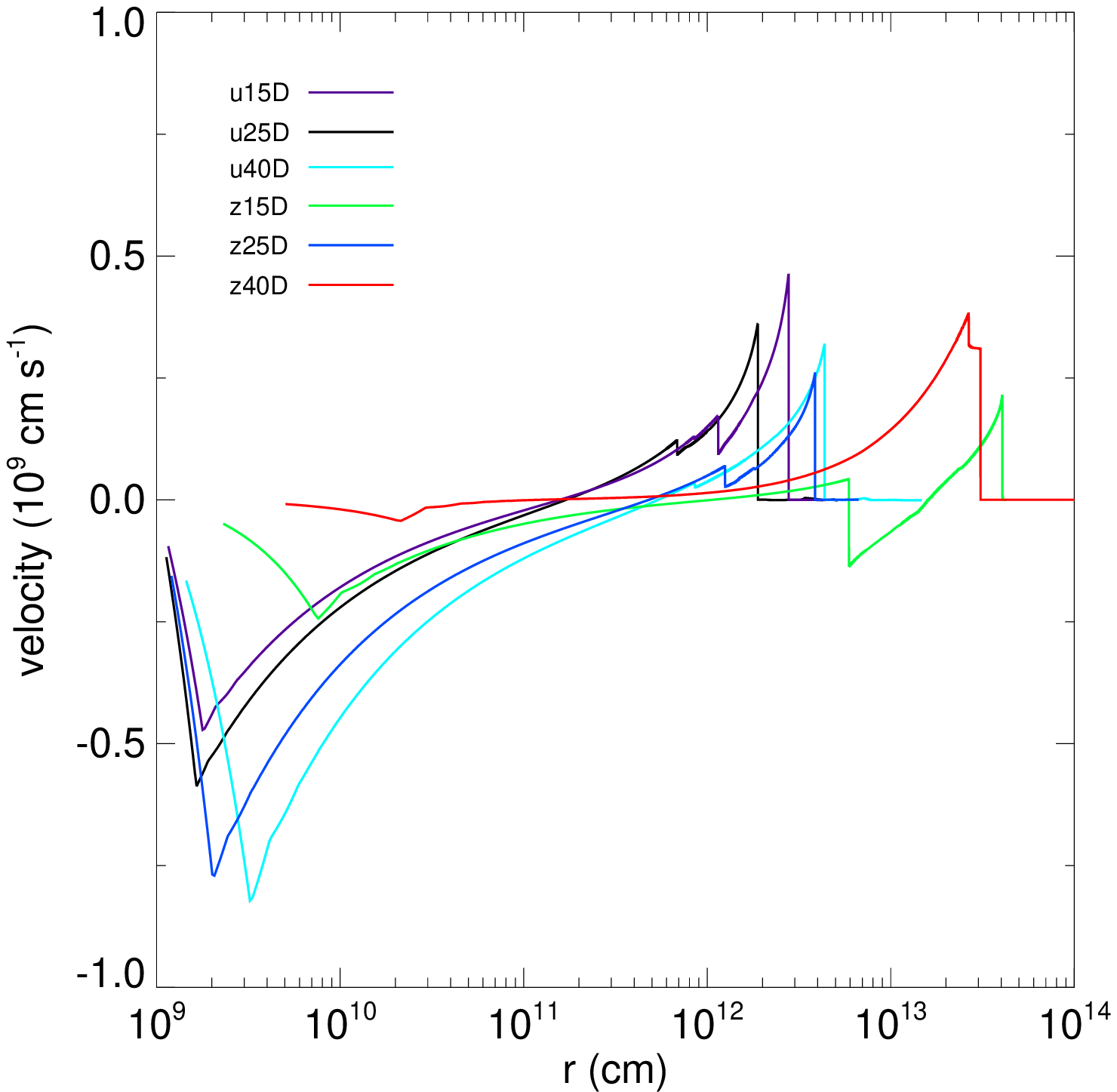,width=0.45\linewidth,clip=} \\
\end{tabular}
\end{center}
\caption{Kepler profiles for the shock, the star and its surrounding envelope for both u-
and z-series Pop III CC SNe as initialized in RAGE.  Left:  densities.  Right:  velocities.}
\label{fig:initprof}
\end{figure*}

We evolve the shock through the outer layers of the star, its surface, and then out into 
the surrounding medium with the radiation hydrodynamics code RAGE \citep{rage}.  
RAGE (Radiation Adaptive Grid Eulerian) is an AMR radiation hydrodynamics code that 
combines a second-order conservative Godunov hydro scheme with grey or multigroup 
flux-limited diffusion to model strongly radiating flows in 1D, 2D, or 3D.  RAGE uses 
atomic opacities compiled from the extensive LANL OPLIB 
database\footnote{http://aphysics2/www.t4.lanl.gov/cgi-bin/opacity/tops.pl}\citep{oplib} 
and can evolve multimaterial flows with several options for the EOS.  We describe the 
physics implemented in our RAGE models in \citet{fet12} (hereafter FET12):  
multispecies advection, 2-temperature (2T) grey flux-limited diffusion, and energy 
deposition due to the radioactive decay of \Ni{} \citep{fet09}.  We note that 2T radiation
transport, in which radiation and matter temperatures are evolved separately, better
models shock breakout and its aftermath than previous 1T models of SN explosions.  
We exclude gravity from our RAGE calculations because JET10 showed that fallback 
onto the central remnant is over before breakout and that its gravity does not strongly 
affect the shock after it reaches the surface of the star.  We evolve mass fractions for 
the same 27 elements as in CASTRO.  

\subsubsection{Model Setup}

We spherically-average 2D densities, velocities, temperatures and mass fractions for the 
explosion, the star, and the circumstellar envelope from CASTRO onto a 50,000 zone 1D 
spherical AMR mesh in RAGE.  Since we do not evolve radiation energy densities (erg 
cm$^{-3}$) in CASTRO, we initialize them in RAGE by assuming that \vspace{0.05in}
\begin{equation}
e_{rad} = aT^4, \vspace{0.05in}
\end{equation}
where $a$ is the radiation constant and $T$ is the gas temperature in CASTRO.  Also, 
we construct the specific internal energy (erg gm$^{-1}$) from $T$ from
\vspace{0.05in}
\begin{equation}
e_{gas} = C_VT, \vspace{0.05in}
\end{equation}
where $C_V = $ 7.919 $\times$ 10$^{7}$ erg K$^{-1}$ is the specific heat of the gas.  
This prescription does not fully hold in the cool outermost layers of red supergiants 
where the temperature falls below $\sim$ 1 eV because it excludes ionization energies
but is adopted there for simplicity.  The initial radius of the shock varies from 4.1 
$\times$ 10$^{13}$ to 1.3 $\times$ 10$^{14}$ cm in the z-series and 2.04 $\times$ 10$
^{12}$ to 4.77 $\times$ 10$^{12}$ cm in the u-series.  Our root grid at setup has a 
resolution of 8.0 $\times$ 10$^8$ cm and an outer boundary at 4.0 $\times$ 10$^{14}$ 
cm for the z-series and a resolution of 1.0 $\times$ 10$^9$ cm and an outer boundary 
at 5.0 $\times$ 10$^{13}$ cm in the u-series.  We allow up to five levels of refinement in 
both the initial interpolation of the profiles onto the setup grid and later throughout the 
simulation. Our setup guarantees that all important features in the profiles are resolved 
by at least 10 grid points and can be subsampled by up to 32 times more points if 
necessary.  We ensure that the explosion is spanned by at least 5000 zones at shock 
breakout so that the photosphere of the shock is fully resolved; failure to do so can lead 
to underestimates of luminosity during post-processing.  

We impose reflecting and outflow boundary conditions on the fluid and radiation flows at 
the inner and outer boundaries of the mesh, respectively.  When the calculation is 
launched, Courant times are initially short due to high temperatures, large velocities and 
small cell sizes.  To speed up the simulation and to accommodate the expansion of the 
ejecta, we periodically resample the profiles onto a larger mesh as the explosion grows.  
Each regrid significantly increases the time step on which the model evolves.  We remap 
just the explosion itself, excluding any medium beyond the shock, and then graft the 
original environment that lies beyond this radius onto the shock on the new grid.  We 
apply the same criteria in choosing a new root grid as in the original problem setup:  any 
important density or velocity structure must be resolved by at least 10 mesh zones and be 
sampled by up to 32 times more zones as needed, and at least 5000 coarse grid zones 
are allocated to the blast after breakout.  The outer boundary of the final largest mesh in 
our models is 5.0 $\times$  10$^{16}$ cm, the greatest distance the ejecta can propagate 
in four months.  

\subsubsection{Circumstellar Envelope}

The ambient media of $z \sim$ 30 and $z \sim$ 10 - 15 Pop III PI SNe are quite different
because the former occur in small cosmological halos and the latter go off in primeval 
galaxies.  Even low-mass Pop III stars can photoevaporate minihalos \citep{wet08a}, and
so at $z \ga$ 20 most die in diffuse relic H II regions with $n \sim$ 0.1 cm$^{-3}$.  Local
densities in $z \sim$ 15 protogalaxies are less well understood but are likely higher.  In 
either case, if the star sheds a wind it will determine the density profile closest to the star.   
When considering PI SNe, \citet{wet12b} allowed for the possibility that massive Pop III 
progenitors have modest winds in addition to being surrounded by a diffuse H II region.  
However, 15 - 40 \Ms\ Pop III stars are less likely to drive strong winds given their 
lower surface temperatures, luminosities, and the fact that stars in this mass range today
manifest only weak winds, even at solar metallicities.  Consequently, in this study we join 
a very low-mass wind profile to the surface of the star:
\vspace{0.05in}
\begin{equation}
\rho_W(r) = \frac{\dot{m}}{4 \pi r^2 v_W}, \vspace{0.05in}
\end{equation}
where $\dot{m}$ is the mass loss rate of the wind and $v_W$ is its speed. The mass loss 
rate is just \vspace{0.05in}
\begin{equation}
\dot{m} = \frac{M_{tot}}{t_{MSL}}, \vspace{0.05in}
\end{equation}
where $M_{tot}$ and $t_{MSL}$ are the total mass loss and main sequence lifetime of the 
star, respectively.  We take $M_{tot}$ to be 0.01 \Ms, $v_W$ to be 1000 km s$^{-1}$, and 
H and He mass fractions to be 76\% and 24\%, respectively.  We extend the wind from the 
surface of the star to the outer boundary of the mesh in RAGE.  We assume that the wind 
has enough time to establish a free-streaming region on the grid and that any bubbles 
blown by the wind are carried far beyond the reach of the ejecta prior to the SN.  Rather 
than calculate the ionization state of the wind \citep{wn06}, we simply assume it to be 
neutral. This is true of stars that die as red supergiants, but not for blue giants, which likely 
ionize their envelopes.  Our u-series light curves and spectra should therefore be taken as 
lower limits.  We show initial density and velocity profiles for our blast models in RAGE in 
Figure \ref{fig:initprof}.

\subsection{SPECTRUM} 

\begin{figure*}
\begin{center}
\begin{tabular}{cc}
\epsfig{file=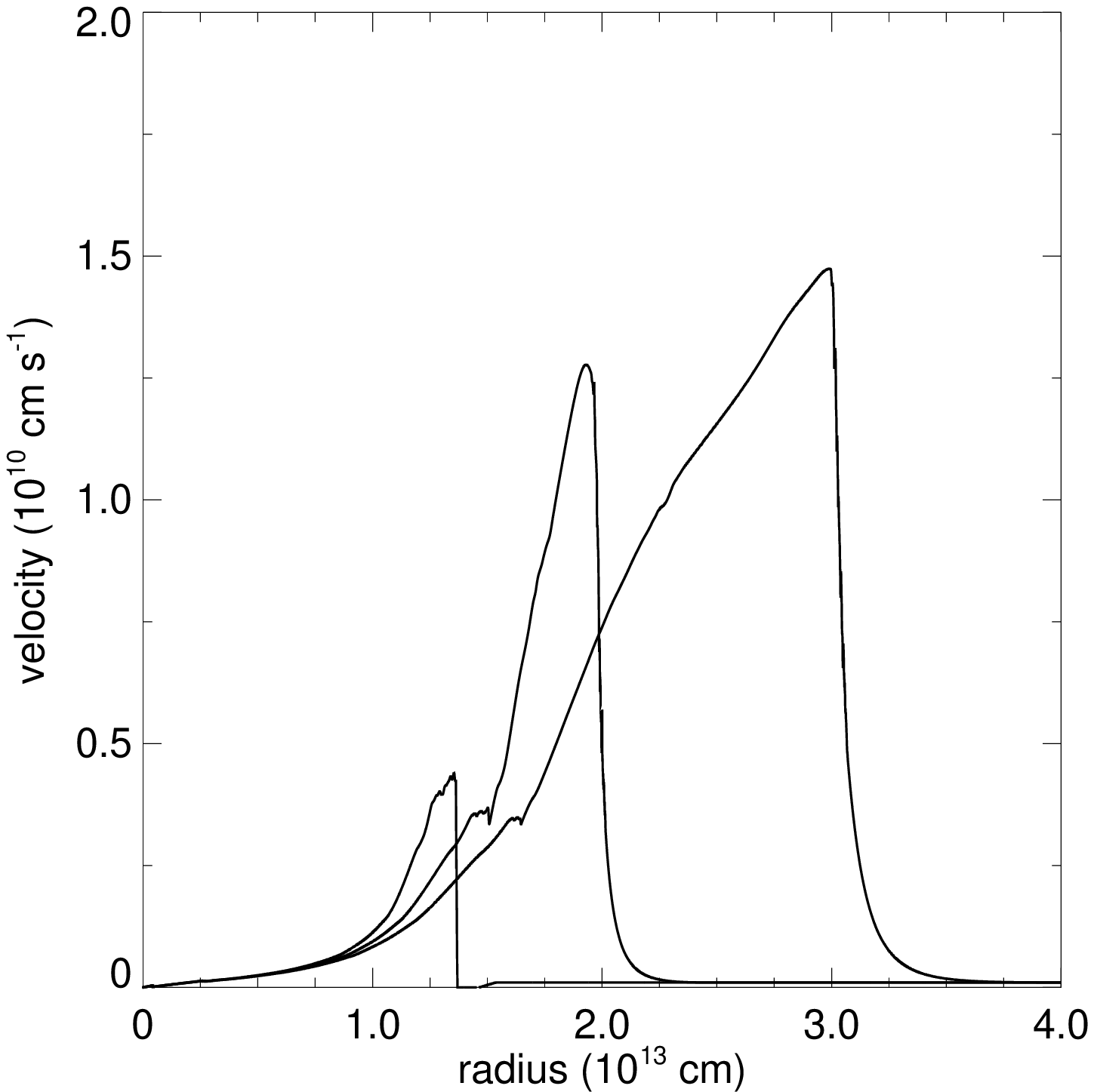,width=0.45\linewidth,clip=} & 
\epsfig{file=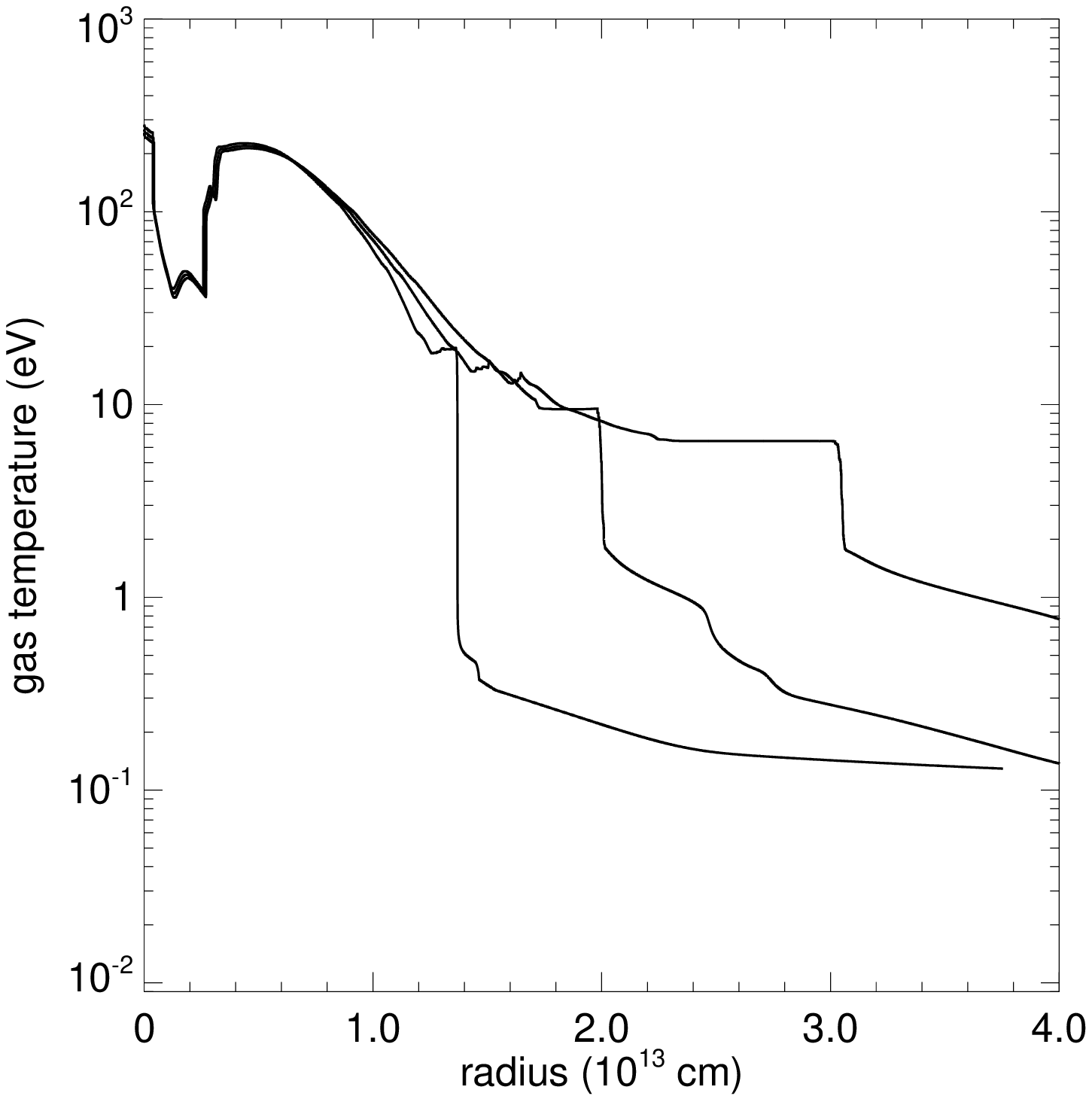,width=0.45\linewidth,clip=} \\
\end{tabular}
\end{center}
\caption{Shock breakout in the u40G (compact blue giant) run. Left panel:  from left to right,
velocities at 1.81 $\times$ 10$^4$ s, 1.87 $\times$ 10$^4$ s and 1.93 $\times$ 10$^4$ s. 
Right panel:  gas temperatures at the same times.}
\label{fig:u40G}
\end{figure*}

\begin{figure*}
\begin{center}
\begin{tabular}{cc}
\epsfig{file=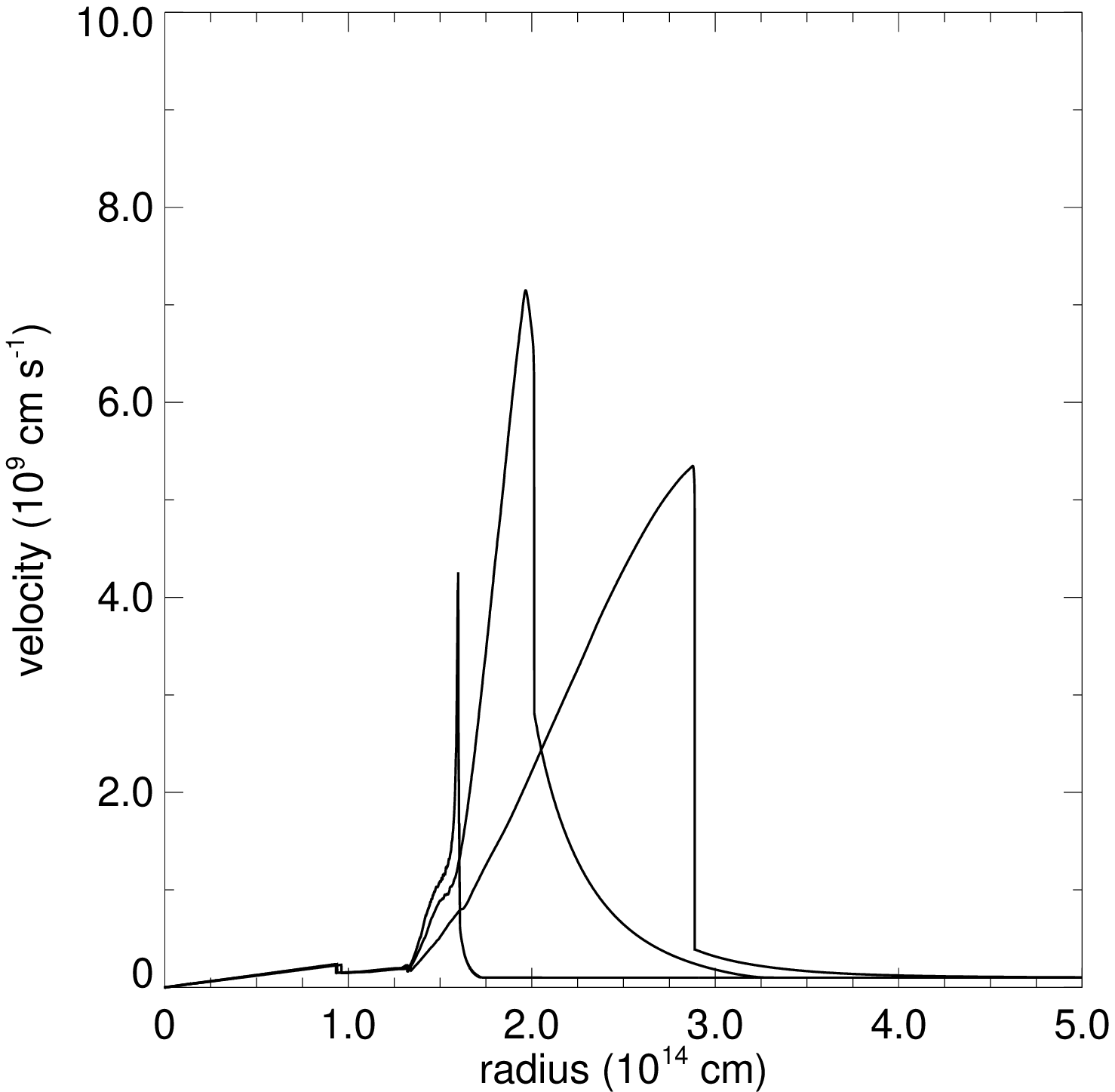,width=0.45\linewidth,clip=} & 
\epsfig{file=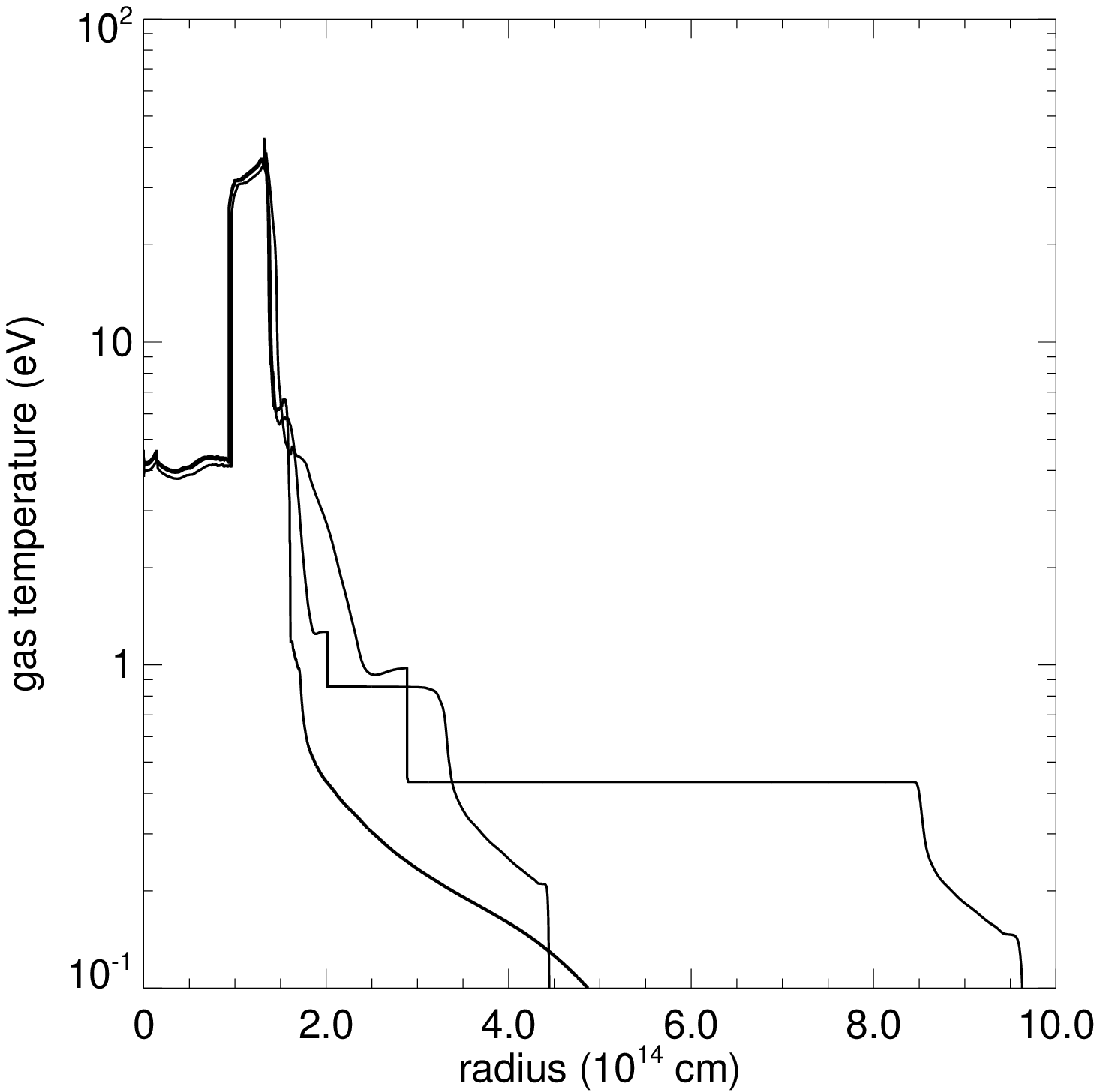,width=0.45\linewidth,clip=} \\
\end{tabular}
\end{center}
\caption{Shock breakout in the z40G (red supergiant) run.  Left panel:  from left to right,
velocities at 3.91 $\times$ 10$^5$ s, 3.97 $\times$ 10$^5$ s and 4.14 $\times$ 10$^5$ s. 
Right panel:  gas temperatures at the same times.}
\label{fig:z40G}
\end{figure*}

\begin{figure*}
\begin{center}
\begin{tabular}{cc}
\epsfig{file=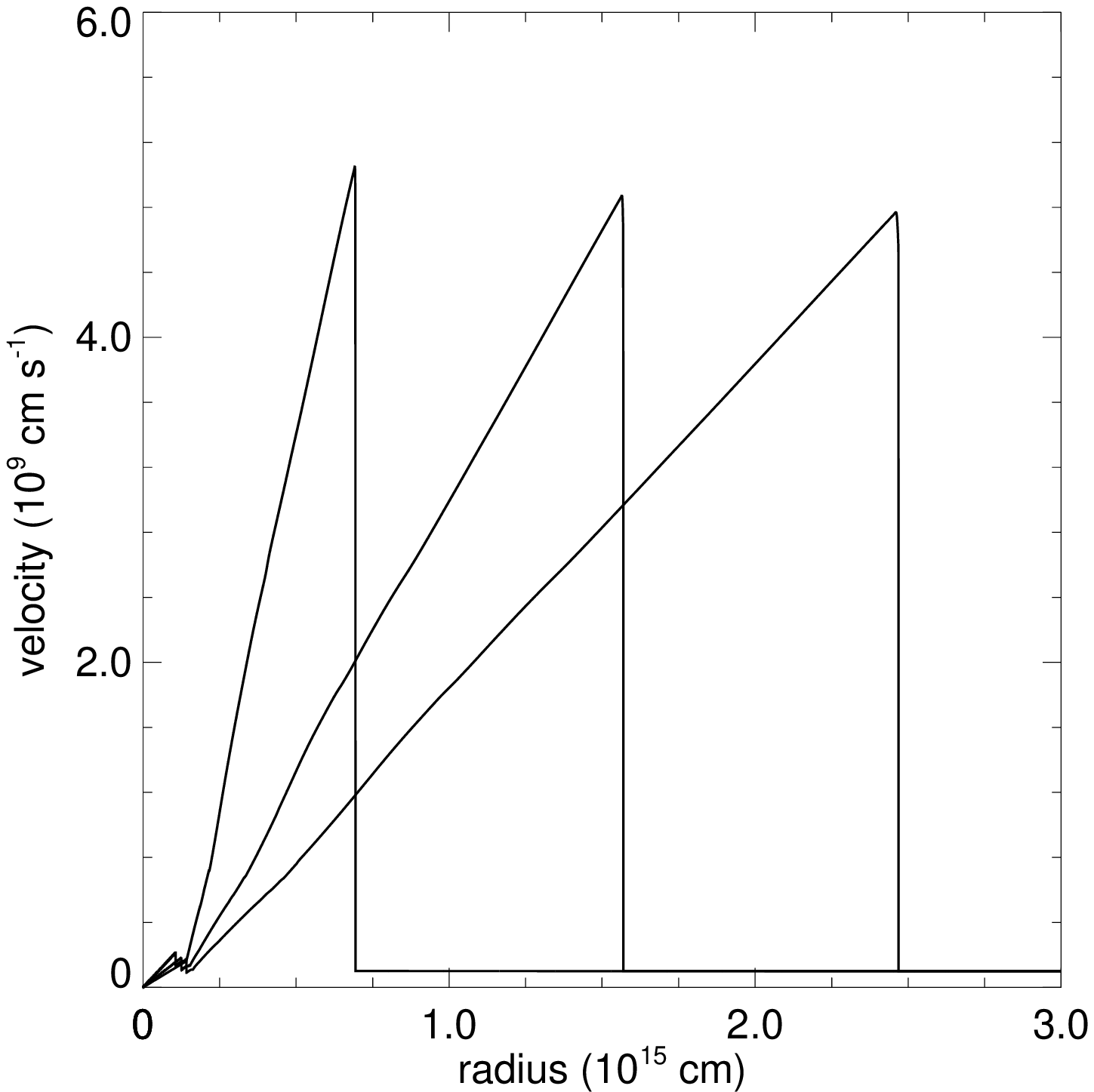,width=0.45\linewidth,clip=} & 
\epsfig{file=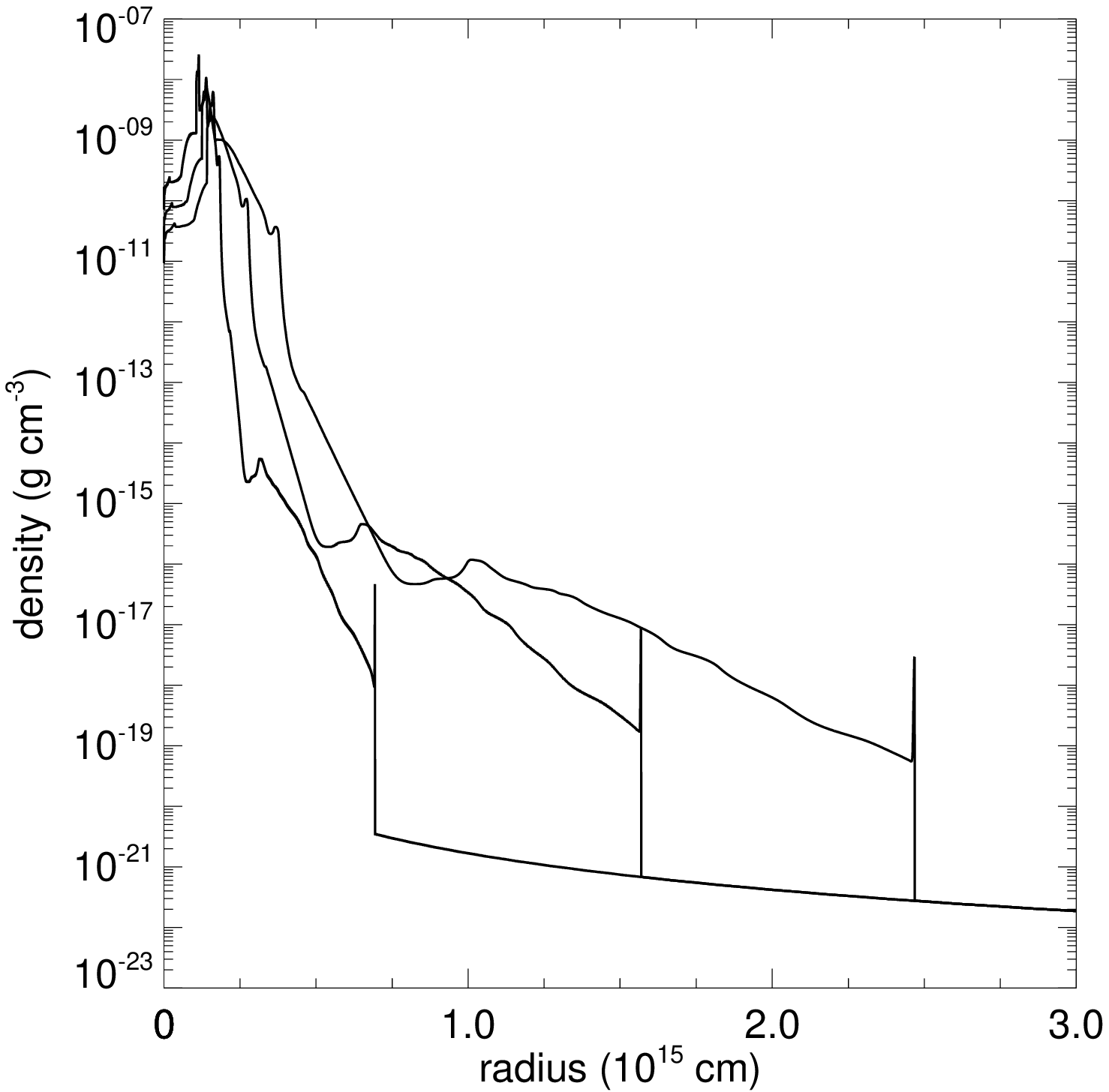,width=0.45\linewidth,clip=} \\
\end{tabular}
\end{center}
\caption{The freely expanding u40G SN at intermediate times. Left panel:  from left to right,
velocities at 4.96 $\times$ 10$^5$ s, 6.80 $\times$ 10$^5$ s and 8.74 $\times$ 10$^5$ s. 
Right panel:  densities at the same times.}
\label{fig:FE}
\end{figure*}

\begin{figure*}
\begin{center}
\begin{tabular}{cc}
\epsfig{file=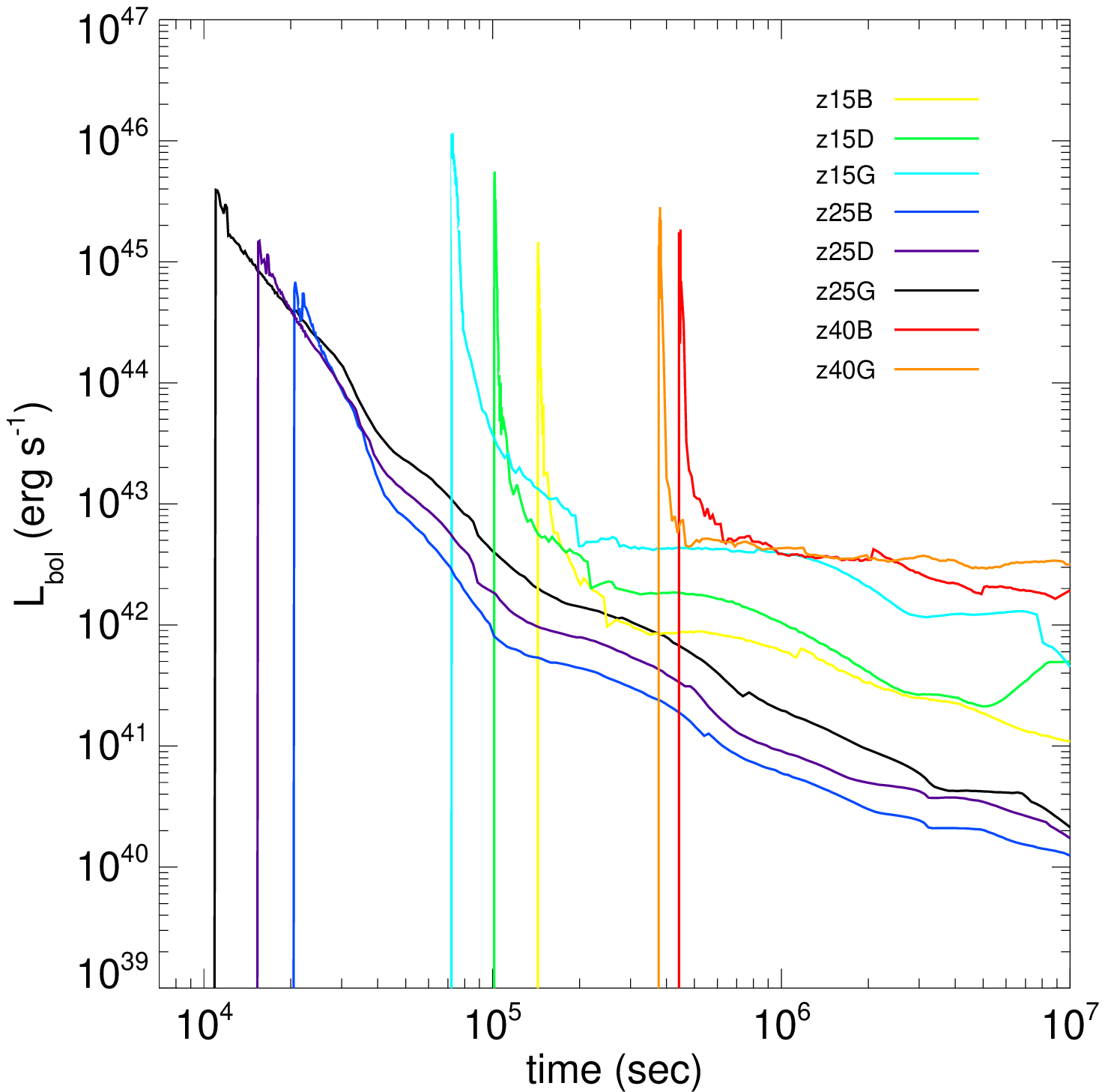,width=0.43\linewidth,clip=} & 
\epsfig{file=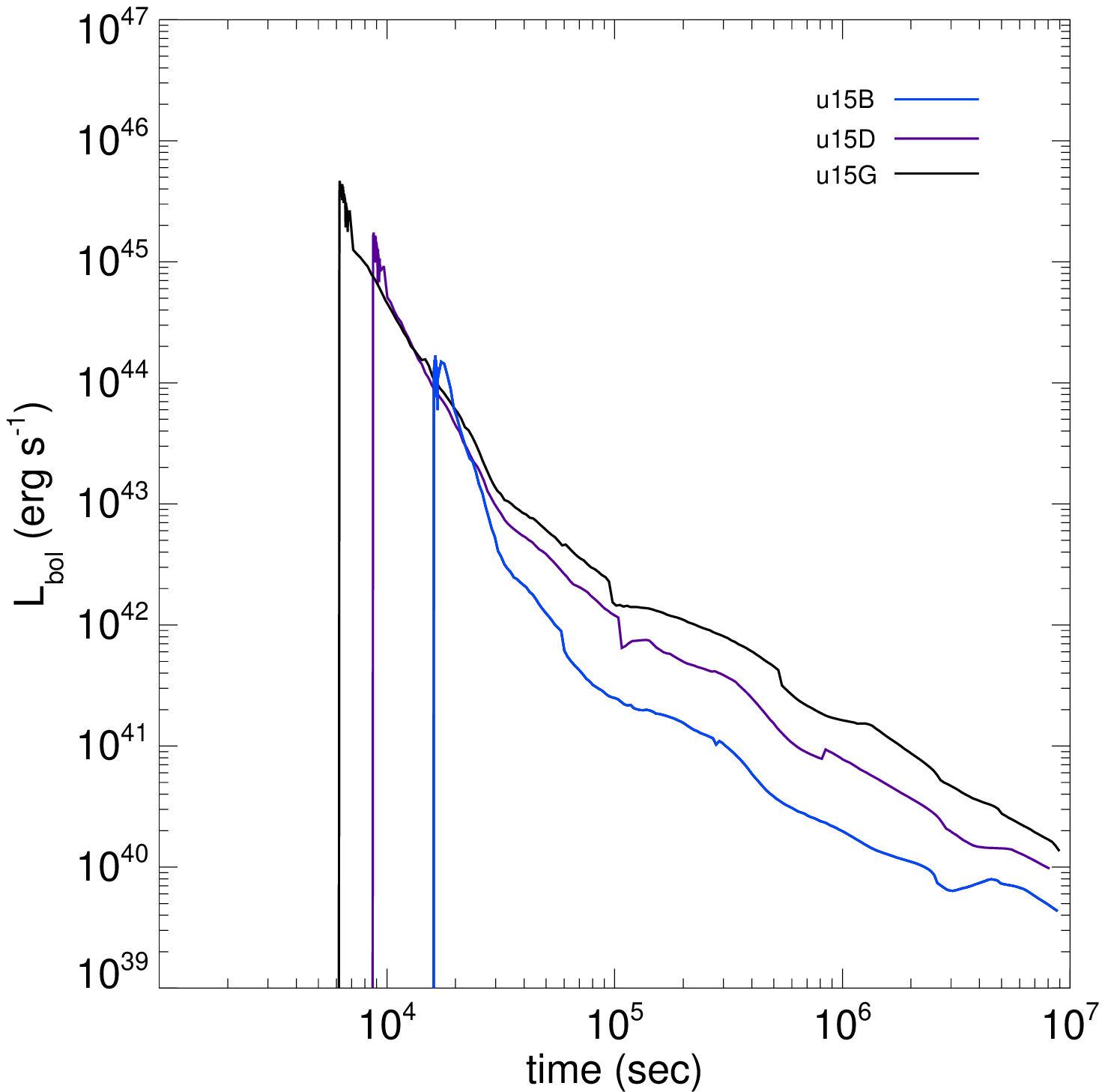,width=0.43\linewidth,clip=} \\
\epsfig{file=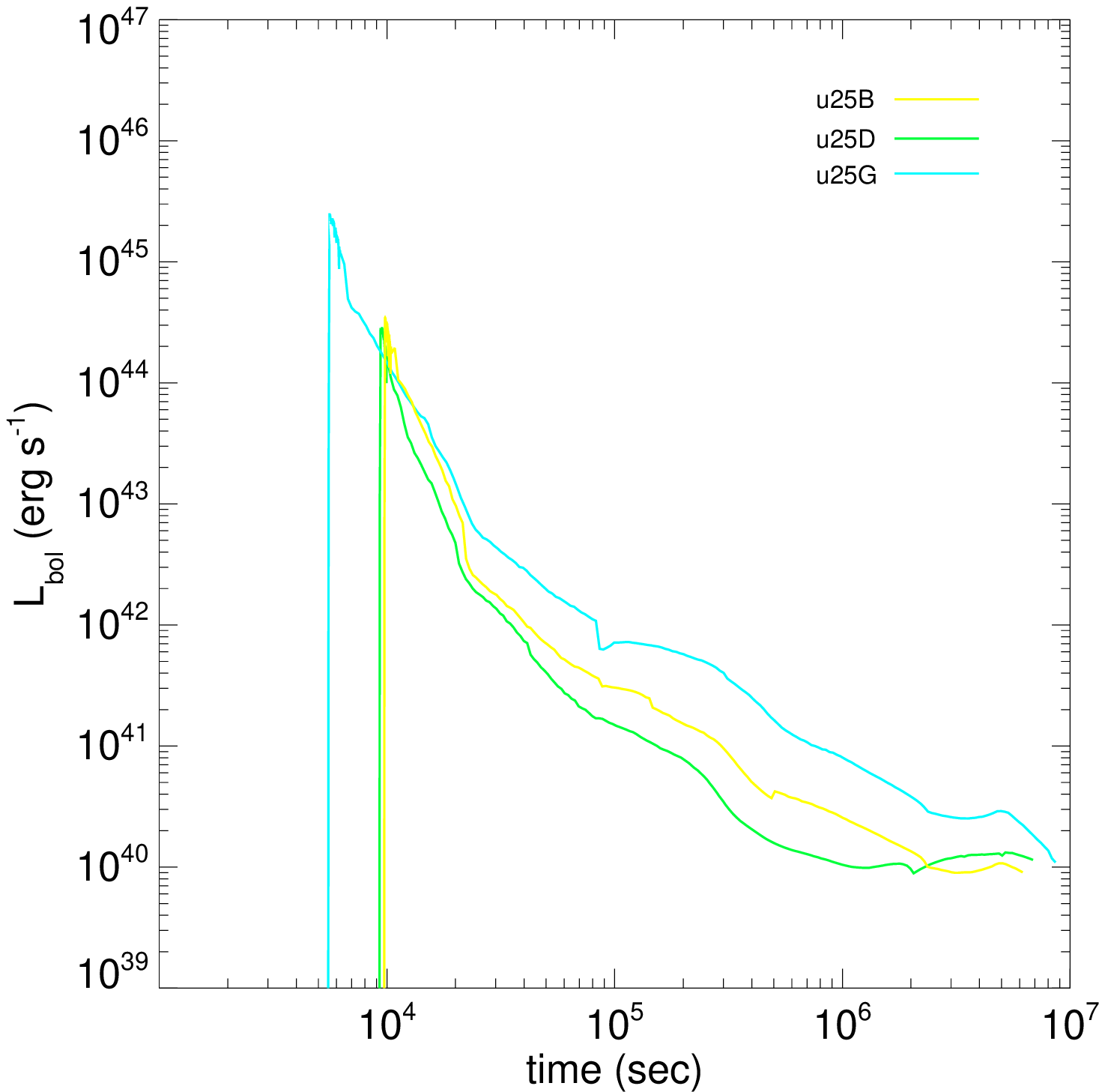,width=0.43\linewidth,clip=} &
\epsfig{file=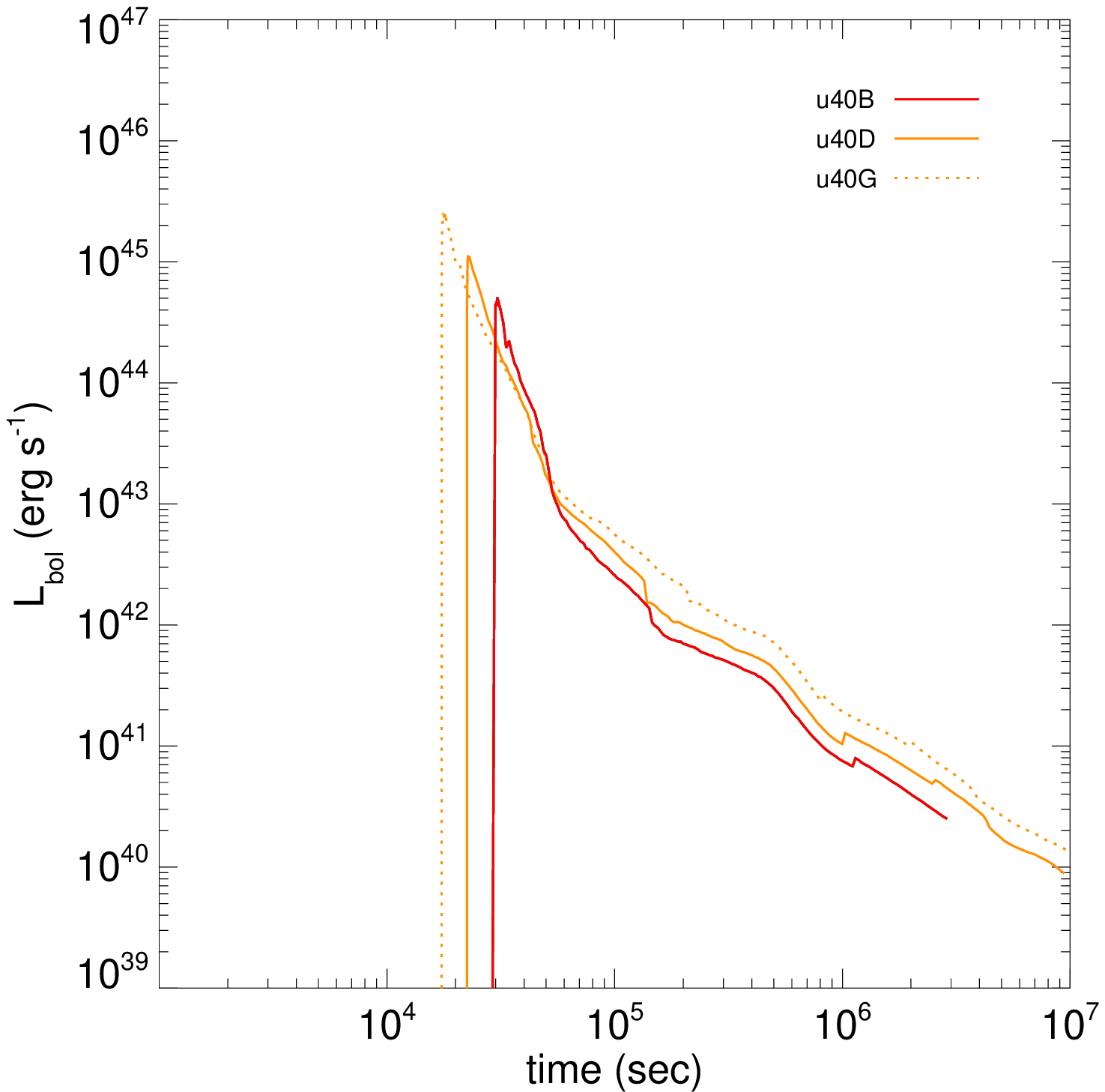,width=0.43\linewidth,clip=}
\end{tabular}
\end{center}
\caption{Bolometric light curves for 17 Pop III CC SNe in the source frame.  Upper left:  
all eight z-series light curves (red supergiant progenitors).  Upper right, lower left and 
lower right are u15, u25, and u40 (blue giant progenitor) light curves, respectively.}
\label{fig:bolLCs}
\end{figure*}

\begin{figure*}
\plottwo{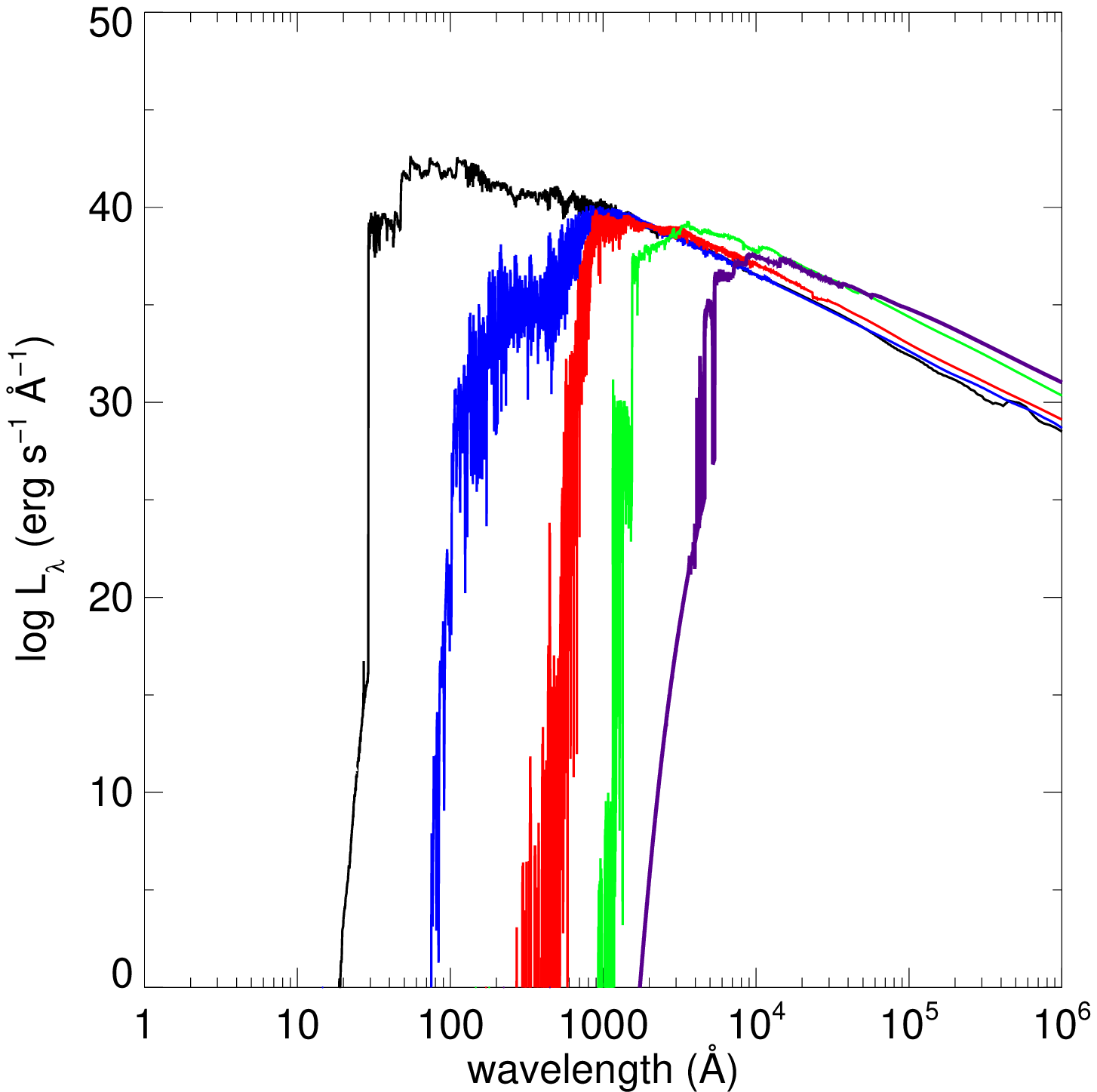}{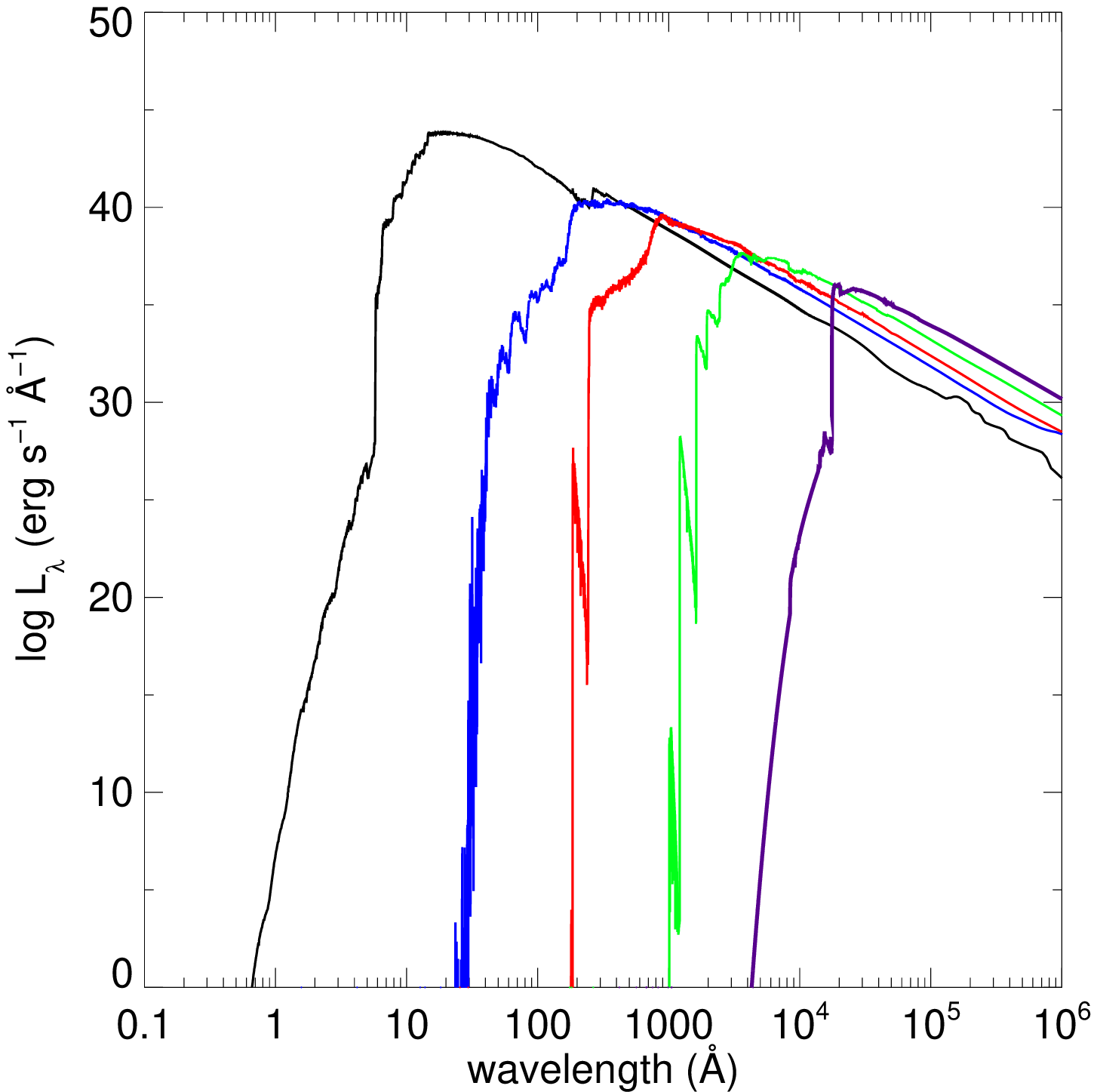}
\caption{Spectral evolution of the fireball.  Left:  z40G at 3.92 $\times$ 10$^5$ s 
(black), 5.55 $\times$ 10$^5$ s (blue), 1.49 $\times$ 10$^6$ s (red), 8.66 $\times$ 
10$^6$ s (green) and 2.95 $\times$ 10$^7$ s (purple).  Right:  u40G at 1.87 $\times$ 
10$^4$ s (black), 6.82 $\times$ 10$^4$ s (blue), 2.06 $\times$ 10$^5$ s (red), 1.21 
$\times$ 10$^6$ s (green) and 9.69 $\times$ 10$^6$ s (purple).\vspace{0.1in}} 
\label{fig:spec}
\end{figure*}

To calculate a spectrum from a RAGE profile, we first map its densities, temperatures, 
velocities and mass fractions onto a 2D grid in the Los Alamos SPECTRUM code.  
SPECTRUM performs a direct sum of the luminosity of every fluid element in this 
discretized profile to compute the total flux emitted by the ejecta along the line of sight 
at every wavelength.  This procedure is described in FET12, and it accounts for Doppler 
shifts and time dilation due to the relativistic expansion of the ejecta.  SPECTRUM also 
calculates the intensities of emission lines and the attenuation of flux along the line of 
sight, thereby capturing both limb darkening and absorption lines imprinted on the flux 
by intervening material in the ejecta and wind.

Gas densities, velocities, mass fractions and radiation temperatures from the finest levels 
of refinement on the RAGE AMR grid are first extracted and ordered by radius into 
separate files, with one variable in each file.  Because of limitations on machine memory 
and time, we map only a subset of these points onto the SPECTRUM grid.  We sample 
the RAGE radiation energy density profile inward from the outer boundary to determine 
the position of the radiation front, where $aT^4$ rises above 1.0 erg/cm$^3$.  The 
maximum depth from which radiation can escape the ejecta is then determined.  For 
radiating shocks this optical depth is often taken to be 
\begin{equation}
\tau \; = \; \frac{c}{v_{sh}}, \label{eqn:taushock}
\end{equation}
where $c$ and $v_{sh}$ are the speed of light and the shock, respectively.  This limit is 
derived from the theory of radiating shocks by assuming that the diffusion limit holds for 
photons in the shock and defining their diffusion timescale through the shock to be 
\citep[e.g.,][]{ohy63}\vspace{0.05in}
\begin{equation}
t_{diff} \; = \; \left(\frac{\Delta x}{\lambda}\right)^2 \frac{\lambda}{c}, \vspace{0.05in}
\end{equation}
where $\Delta x$ is roughly the width of the shock and $\lambda$ is the photon mean 
free path.  Setting the diffusion velocity of the photons through the shock to be 
\vspace{0.05in}
\begin{equation}
v_{diff} \; = \; \frac{\Delta x}{t_{diff}} \vspace{0.05in}
\end{equation}
and recalling that \vspace{0.05in}
\begin{equation}
\tau \; = \; \frac{\Delta x}{\lambda} \vspace{0.05in}
\end{equation}
it follows that \vspace{0.05in}
\begin{equation}
v_{diff} \; = \; \frac{c}{\tau}. \vspace{0.05in}
\end{equation}
If one equates the photon diffusion velocity with $v_{sh}$, it follows that the greatest 
optical depth from which photons can escape the shock is given by equation 
\ref{eqn:taushock}.  This derivation also assumes that the opacity is constant.  For 
the 50,000 km s$^{-1}$ flows typical of SNe, $\tau$ $\sim$ 6.

However, the assumption of the diffusion limit and constant opacity likely does not 
fully hold, particularly during shock breakout when photons are no longer trapped by
the shock, so we adopt $\tau = $ 20 as a more careful limit.  We find the radius of the 
$\tau = $ 20 surface in the ejecta by integrating the optical depth due to Thomson 
scattering inward from the outer boundary, taking $\kappa_{Th}$ to be 0.288 for H and 
He gas at primordial composition.  

The extracted gas densities, velocities, temperatures and species mass fractions are 
then interpolated onto a 2D grid in $r$ and $\theta$ in SPECTRUM whose inner 
boundary is that of the RAGE mesh and whose outer boundary is 10$^{18}$ cm.  One 
hundred uniform zones in log radius are assigned from the center of the grid to the 
$\tau =$ 20 surface, and the region between the $\tau =$ 20 surface and the edge of 
the radiation front is partitioned into 5000 uniform zones in radius.  The wind between 
the front and the outer edge of the grid is divided into one hundred uniform zones in log 
radius, for a total of 5200 radial bins.  The data within each of these new radial bins is 
mass-averaged to guarantee that the SPECTRUM grid captures very sharp features 
from the original RAGE profile.  The grid is uniformly divided into 160 bins in $\mu =$ 
cos$\,\theta$ from -1 to 1.  Our mesh fully resolves regions of the flow from which 
photons can escape the ejecta and only lightly samples those from which they cannot.

The sum of the luminosities over all wavelengths in one spectrum yields the bolometric 
luminosity of the SN at that moment.  Many such luminosities computed over a range 
of times constitutes the light curve of the explosion.  We cover shock breakout with 50 
spectra uniformly spaced in time and the rest of the light curve with 200 spectra that are
logarithmically distributed in time out to 4 months. Each SPECTRUM calculation requires 
3 - 6 hours on 32 processors on LANL platforms.

\section{Blast Profiles, Light Curves and Spectra}

Hydrodynamical profiles for shock breakout are shown for the fiducial cases u40G, a 
blue giant progenitor, and z40G, a red supergiant progenitor, in Figures \ref{fig:u40G} 
and \ref{fig:z40G}. The reverse shock in the z40G velocity profile in Figure \ref{fig:initprof}, 
which vigorously mixes the interior of the star prior to breakout \citep[Figure 5 of][]{jet09b}, 
is gone by the time the shock reaches the surface of the star at 3.91 $\times$ 10$^5$ s.  
With no reverse shock there is no further Rayleigh-Taylor mixing, and the metals are 
essentially frozen in mass coordinate as the ejecta expands.  Our procedure for spherically
averaging 2D CASTRO mass fractions to approximate their radial distribution in 1D in RAGE
therefore captures all mixing in the explosion.  The SN is not visible to an external observer 
before breakout because electrons between the surface and the shock scatter the photons. 

When the shock breaks through the surface of the star its velocity abruptly increases in 
the steep density drop there.  The sudden acceleration heats the shock, and as photons 
stream free from it they blow off the outer layers of the star at high velocities.  This effect
is far stronger in u40G, where the outermost wispy layers of the star are accelerated to 
nearly half the speed of light.  The gas velocity only doubles in z40G because the density
of the wind envelope is much lower at the surface of the star and because the shock is 
much cooler at breakout.  Since RAGE does not employ relativistic hydrodynamics there 
is some inaccuracy in our peak breakout velocities for the u-series SNe.  However, as we 
shall show, shock breakout is irrelevant to the detection of these events at very high 
redshift because its x-rays and hard UV are all absorbed by the neutral IGM, so this is not 
a serious limitation.  

The breakout transient is visible as the flat plateau in gas temperature that moves ahead 
of the ejecta.  This plateau traces the leading edge of the radiation front. The height of the 
plateau is the temperature to which the radiation pulse heats the gas as it passes through 
it, not the temperature of the fireball itself, which is much higher.  The u40G shock is much 
hotter than the z40G shock at breakout ($\sim$ 100 eV vs. 10 eV) because it has broken 
out of a much more compact star and done less $PdV$ work on its surroundings.  As a 
result, the radiation front initially heats the wind to 20 eV in u40G but to only 1 eV in z40G.  
As the shock expands its spectrum softens, and the temperature of the gas behind the 
radiation front falls over time:  from 20 eV to 8 eV in u40G from 1.81 $\times$ 10$^4$ s to 
1.93 $\times$ 10$^4$ s and from 1.2 eV to 0.4 eV in z40G from 3.91 $\times$ 10$^5$ s to 
4.14 $\times$ 10$^5$ s.

In both explosions, shock breakout coincides with radiation breakout because the moderate 
wind density at the surface of the star cannot confine the radiation front.  As both shocks 
descend the $r^{-2}$ wind gradient, the u40G SN accelerates but the z40G SN speeds up 
and then decelerates.  The u40G shock continues to accelerate even though ambient 
densities are 1000 times higher at the surface of the star because the explosion is much 
hotter and the shock is more radiatively driven than the much cooler z40G shock.  Soon 
after breakout, both SNe assume self-similar free expansion velocity profiles and 
homologously expand, as we show in Figure \ref{fig:FE} for u40G.  This self-similar 
expansion continues until the end of the simulations at 4 months.  Eventually, spherical 
dilution of the ejecta renders the SN transparent, and it enters the nebular phase.  At this 
stage, fluid elements in the ejecta may no longer remain in radiation equilibrium with their 
neighbors so the true opacity may deviate from the OPLIB opacities, which assume local 
thermodynamic equilibrium (LTE).  Also, the Kirchhoff-Planck relation \citep[equation 4 
of][]{fet12} may not fully hold in the SPECTRUM code.  There may therefore be some 
inaccuracy in our spectra at late times, but these events are only visible in the NIR at much 
earlier stages of the explosion, as we later show.

We note that early simulations of shock breakout exhibited a spike in the matter temperature 
ahead of the shock \citep{Ens92}.  Our simulations assume that the ions and electrons are 
strongly coupled, leading to only mild deviations between the matter and radiation temperature.  
As such, we do not expect, and our simulations do not produce, a strong spike in the matter 
temperature at the front of the shock, in agreement with many recent results \citep{tomin11,
moriya12,tbn13}.  Since \citet{Ens92} also assume electron/ion coupling, one would not expect 
spike in temperature in their simulations either.  The temperature spike in their calculations 
has been attributed to the simplification they use to calculate their opacities \citep{tbn13}.

However, as the ions and electrons decouple, the ions no longer lose energy to the radiation 
field and the ion temperature can dramatically spike at the shock front.  As the shock density 
decreases (e.g., in a supernova remnant), the ions and electrons no longer couple, and our
"strongly-coupled" assumption breaks down.  Depending on the density of the surroundings, 
this decoupling can occur during the supernova outburst, leading to increased X-ray emission.  
This may be the cause of the spike in the \citet{kl78} calculations.

\subsection{Light Curves / Spectra}

We show bolometric luminosities for our Pop III CC SNe in Figure \ref{fig:bolLCs}.  
Peak luminosities at shock breakout vary from 8 $\times$ 10$^{44}$ to 1.5 $\times$ 10$^{
46}$ erg s$^{-1}$ in the z-series and from 3 $\times$ 10$^{44}$ to 5 $\times$ 10$^{45}$ erg 
s$^{-1}$ in the u-series.  At early times the light curve is powered primarily by the conversion 
of kinetic energy into thermal energy by the shock, so for a given progenitor mass the peak 
luminosities increase with explosion energy.  They peak at earlier times with greater $E$ 
because the shock reaches the surface of the star in less time.  Breakout also generally 
happens later with more massive stars because they have greater radii in both series, with 
the exception of the 25 \Ms\ star because it dies with the smallest radius, as shown in Figure 
\ref{fig:initprof}.  For a given progenitor mass and $E$, u-series SNe are somewhat less 
luminous than z-series SNe, and this is due to a tradeoff between shock temperature at 
breakout and the radius of the star:  \vspace{0.075in}
\begin{equation}
L = 4 \pi \epsilon r^2 \sigma T^4. \label{eq:BB} \vspace{0.075in}
\end{equation} 
Here, $\sigma$ is the Stefan-Boltzmann constant, $\epsilon$ is the greybody correction 
to the blackbody luminosity assumed for the shock ($\epsilon \sim$ 0.1), and $T$ is the 
temperature of the shock at the $\tau_{Th} = $ 1 surface, where $\kappa_{Th} = $ 0.288.
The red z-series stars have radii 10 - 30 times greater than u-series stars of equal mass
but their shocks have lower temperatures at breakout because they must do more $PdV$
work before crashing through the surface of the star.  Their respective temperatures are
evident in their spectra at breakout, which we show in Figure \ref{fig:spec}.  The u40G
model has much more spectral energy below 10 \AA\ than z40G at breakout because the 
fireball has a much higher temperature. The duration of the breakout transient is governed 
by the light-crossing time of the star and is 10 - 30 times greater in the z-series than in the 
u-series because of their larger radii.

As the shock expands it cools, and its emission at later times is powered by the radioactive
decay of \Ni.  The duration of this emission can be approximated by the radiation diffusion
timescales in the ejecta:  \vspace{0.075in}
\begin{equation}
t_d \sim \kappa^{\frac{1}{2}} {M_{ej}}^{\frac{3}{4}} E^{-\frac{3}{4}}, \vspace{0.075in}
\end{equation}
where $M_{ej}$ is the mass of the ejecta, $\kappa$ is the average opacity of the ejecta, 
and $E$ is the explosion energy.  The u-series SNe dim after about 3 months but the 
z-series explosions exhibit much more extended emission reminiscent of Type IIp SNe,
whose progenitors are also thought to be red supergiants.  In particular, the z40 series
remain above 10$^{42}$ erg s$^{-1}$ for up to 300 days.  However, because the shock
cools over time, the region of its spectrum that is redshifted into the NIR in the observer 
frame dims below detection limits well before 1 year, as we discuss below.  

We show the evolution of the z40G and u40G spectra in Figure \ref{fig:spec} \citep[compare 
to Figure 2 in][]{tomin11}. Unlike with much more energetic Pop III PI SNe, the outer regions 
of the ejecta and the envelope are never fully ionized so bound-bound and bound-free 
transitions absorb most of the flux at the short wavelength limit of the spectrum. As the shock 
expands and cools its surrounding envelope removes more flux at high energies in the 
spectrum and continuum emission at long wavelengths slowly rises.  The flux at longer 
wavelengths increases with time because the ejecta cools and its surface area grows.  From 
these profiles, it is clear that fitting a blackbody profile to a bolometric luminosity in order to 
approximate a spectrum can lead to serious overestimates of flux at short wavelengths 
because it erroneously reinstates luminosity that is actually removed by the envelope.  This 
caveat is especially pertinent to detection thresholds for high-redshift SNe because flux from 
this region of the spectrum is a principal component of the NIR signal of the explosion in the 
observer frame.  The collision of the shock with a realistic circumstellar envelope (which sets 
its temperature) and the opacity of the ejecta and envelope also crucially shape the spectra 
in the source frame, and thus the NIR light curve in the observer frame. 

\section{Pop III CC SN Detection Thresholds}

\begin{figure*}
\plottwo{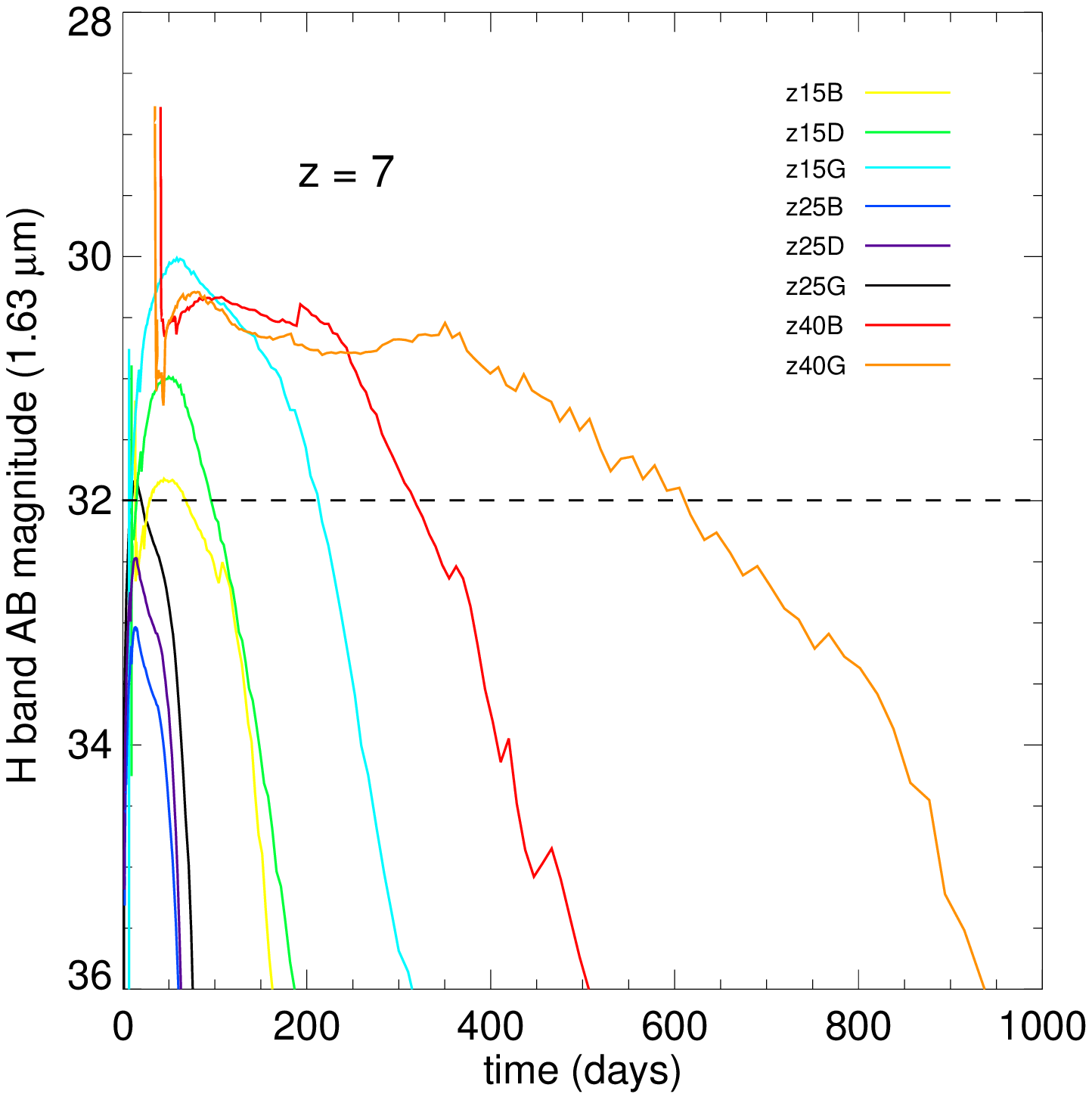}{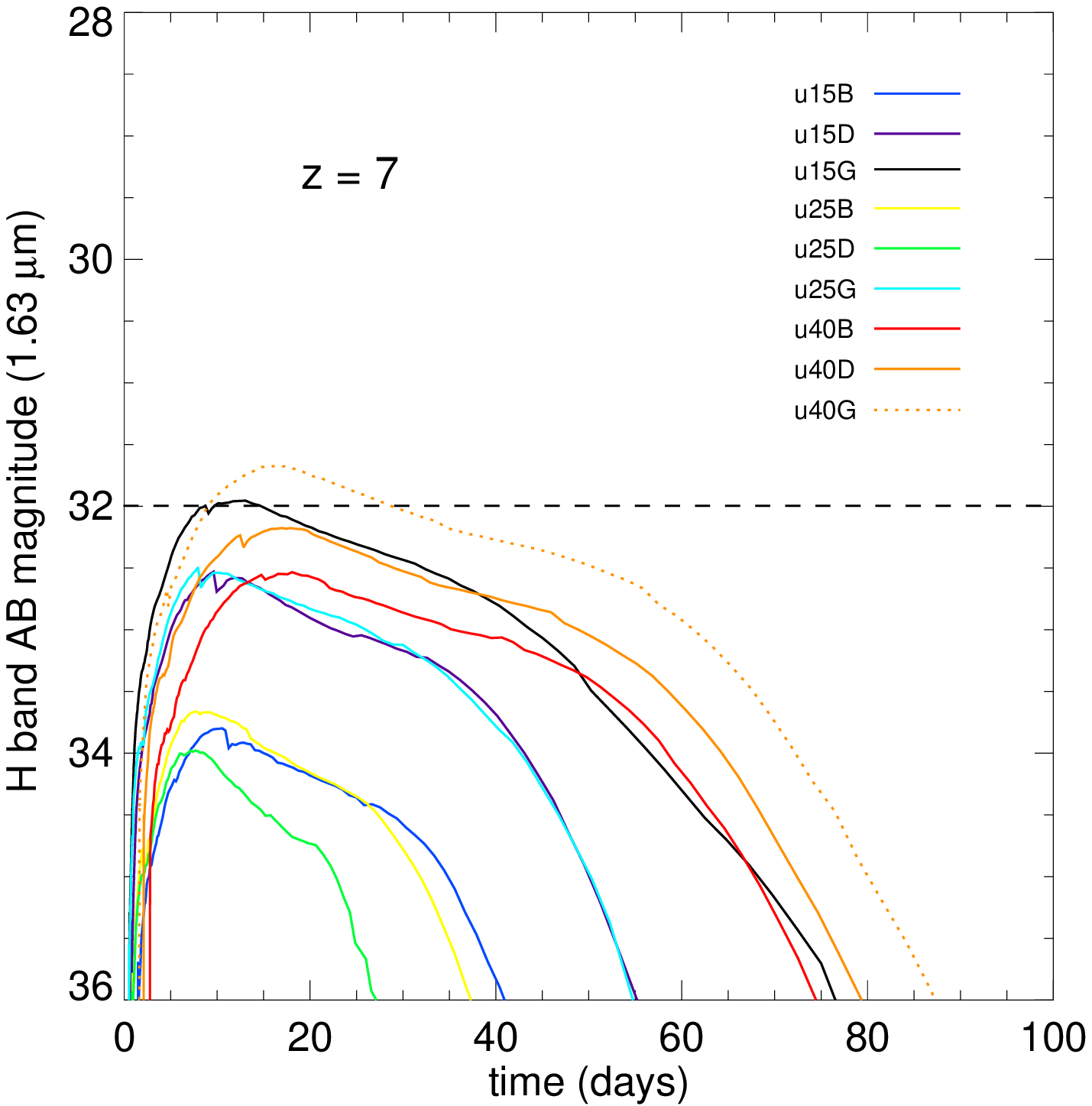}
\caption{NIR light curves for $z =$ 7 Pop III CC SNe at 1.63 $\mu$m in the observer frame.
Left:  z-series.  Right:  u-series}
\label{fig:nirz7}
\end{figure*}

We calculate NIR light curves for our SNe with the photometry code developed by \citet{su11}.  
Each spectrum is redshifted before removing the flux that is absorbed by intervening neutral 
hydrogen along the line of sight according to the method of \citet{madau95}. The spectrum is 
then dimmed by the required cosmological factors.  We linearly interpolate the least sampled 
data between the input spectrum and filter curve to model the light curve in a given filter.  

\subsection{NIR Light Curves}

\begin{figure*}
\plottwo{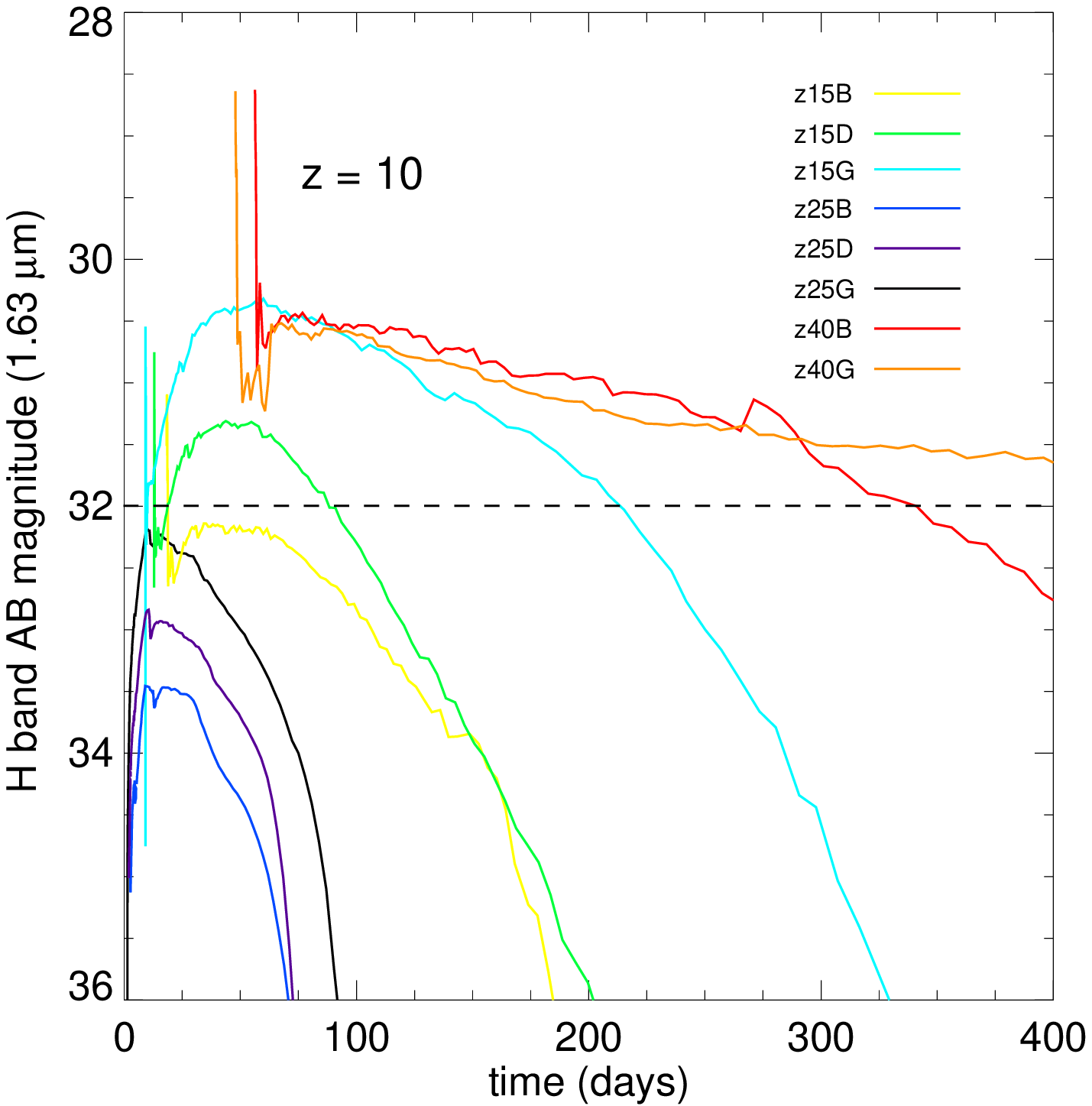}{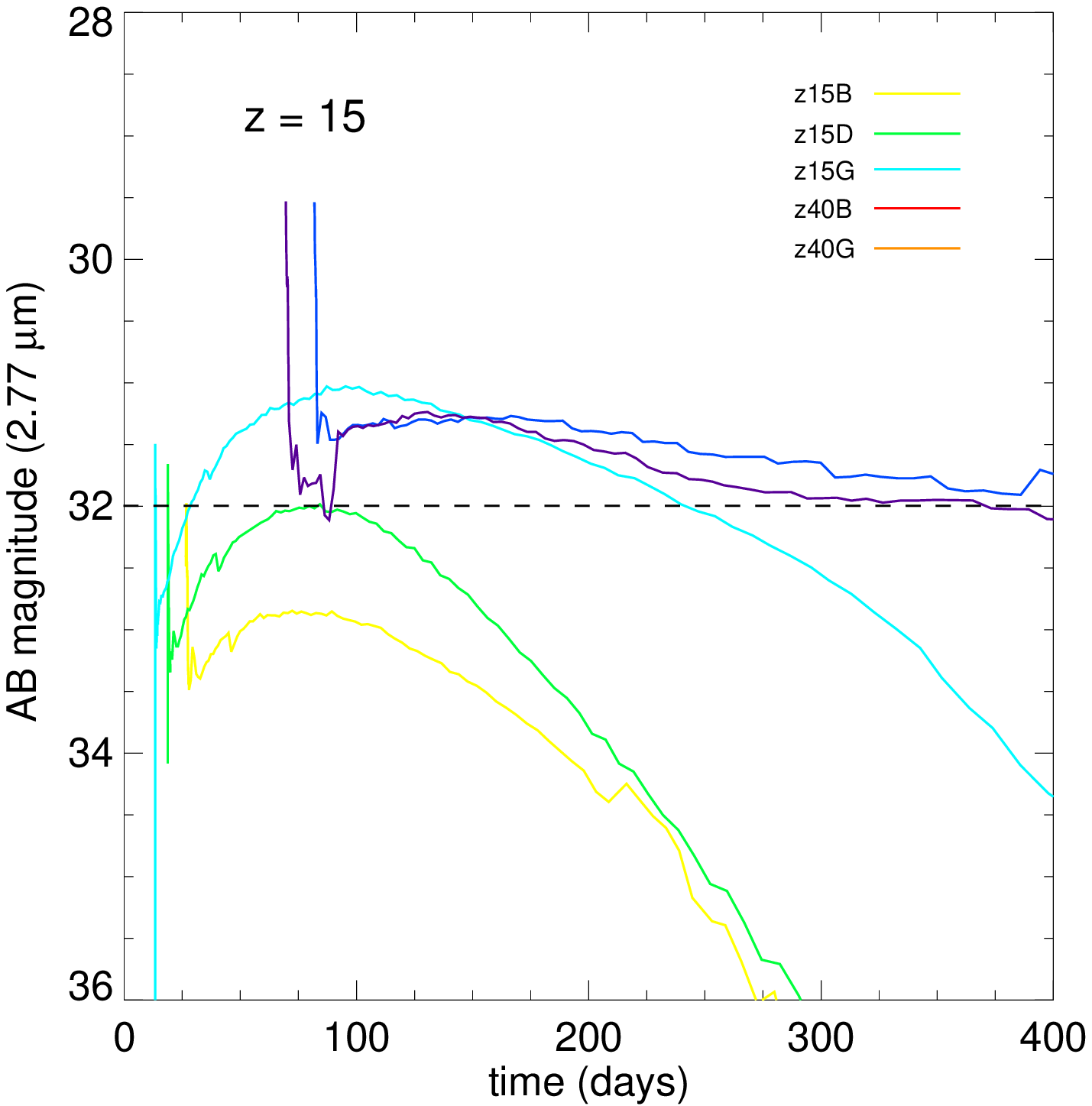}
\caption{NIR light curves for z-series SNe at 1.63 $\mu$m (left, $z =$ 10) and at 2.77 $\mu$m 
(right, $z =$ 15) in the observer frame.
\vspace{0.1in}}
\label{fig:NIRz10}
\end{figure*}

At every redshift for each supernova, we calculate the NIR signal in \textit{JWST} NIRCam 
filters above and below the Lyman limit to find the filter in which the SN is brightest.  As in 
\citet{wet12a}, we find that the CC SNe in our study are most luminous just redward of 1216 
\AA\ in the source frame at the redshifts we consider.  We show NIR light curves for z-series 
and u-series SNe at $z =$ 7, 10 and 15 in Figures \ref{fig:nirz7} and \ref{fig:NIRz10}.  The 
NIRCam photometry limit is AB magnitude 32, and so six of the eight z-series SNe will be 
visible for 20 - 600 days at $z =$ 7 but only one of the u-series explosions, u40G, will be 
visible, and only for $\sim$ 30 days.  Five of the z-series SNe will be visible at $z =$ 10:  
z15B for less than 1 day, z15D for 75 days, z15G for 220 days, z40B for 275 days and z40G 
for over 400 days.  At $z =$ 15 the z15G SN is visible for 250 days and the z40B and z40G 
explosions can be seen for 300 days.  The peak in emission for all the SNe advances to later 
times and is broader at earlier epochs because of cosmological redshifting. For a given mass, 
the NIR luminosity increases with the explosion energy.

Shock 
breakout is not visible from earth in any of these SNe because the x-rays and hard UV in the 
transient are absorbed by neutral H at high redshift.  The flux that is redshifted into the NIR 
varies much more rapidly (on timescales of $\sim$ 100 days) than the bolometric luminosity (3 
- 7 years) in the observer frame because the spectra evolve as the fireball expands.  The light 
curves rise more rapidly than they fall, so they are most recognizable as transients in their early 
stages. However, given \textit{JWST} survey times of 1 - 5 years, it will be possible to identify 
these events as SNe at any stage above detection threshold.  Because they are much dimmer 
than Pop III PI SNe and because $z \sim$ 7 galaxies will be more luminous than $z \sim$ 10 - 
15 protogalaxies, some Pop III CC SNe may be somewhat more challenging to discriminate 
from their host galaxies (with which they likely overlap in color-color space). However, they will 
still be more luminous than most galaxies of that epoch and may be much more frequent than 
PI SNe if baryons in some metal-free halos collapse and fragment into small swarms of stars 
rather than a few very massive ones, as the latest numerical simulations suggest.

The prospects for detecting Pop III CC SNe with \textit{JWST} at $z \ga$ 10 are better than 
those for PI SNe \citep[e.g.,][]{hum12} because of their higher rates, but they will be too dim 
to be found in future all-sky NIR surveys such as \textit{Euclid}, whose detection limit is AB 
magnitude 24 at 2 $\mu$m, or \textit{WFIRST} and \textit{WISH}, whose photometry limits 
will be AB magnitude 27 at 2.2 $\mu$m.  They also will fall below the detection threshold in 
the Y-band centered at 1.0 $\mu$m for the \textit{Large Synoptic Survey Telescope} 
(\textit{LSST}) and the \textit{Panoramic Survey Telescope \& Rapid Response System} 
(\textit{Pan-STARRS}), whose photometry limits will be at most AB magnitude 25 and 27, 
respectively.  Our SPECTRUM calculations show that these SNe will be even dimmer in the 
optical in \textit{LSST} and \textit{Pan-STARRS} because of extinction by the IGM at high 
redshift.  We note that while we have considered only Pop III CC SNe, our detection limits 
hold for progenitors of any metallicity.  The central engine of the explosion depends mostly
on the entropy profile of the inner 3 - 4 \Ms\ of the star, which does not vary strongly with 
metallicity \citep{cl04,wh07} \citep[see also Figure 1 in][]{wf12}.  The light curves and spectra 
of CC SNe at higher metallicities should therefore be bracketed by those of our compact blue 
giant and red supergiant Pop III progenitor stars.

\section{Conclusion}

Although direct detections of Pop III CC SNe will not probe the very first stellar populations, 
they will reveal the properties of stars in $z \sim$ 10 - 15 galaxies and trace their formation 
rates and early galactic chemical evolution in general \citep{wa05,wl05,oet05,tfs07,wet08b,
tss09,wet10,get10,maio11,hum12,jdk12,wise12}.  Furthermore, because they will likely be 
brighter than most primitive galaxies of that era, explosions of low-mass Pop III stars (and 
indeed any CC explosion) will reveal the existence of primeval galaxies on the sky that 
might otherwise escape detection by \textit{JWST} or future ground-based 30-meter class 
telescopes.  We note that while we have considered explosion energies of 0.6 - 2.4 Be, 
higher energies may be possible for some CC SNe and would be visible at even higher 
redshifts.

There are also scenarios in which CC explosions can produce luminosities that rival those of 
PI explosions that could be visible at much earlier epochs.  For example, if the shock collides 
with a dense shell ejected during a luminous blue variable (LBV) outburst prior to the SN, an 
extremely bright event in the UV can result that could be detected above z $\sim$ 15 
\citep[e.g.][]{nsmith07a,nsmith07b,vmarle10,moriya10,moriya12,tet12}. We are now studying 
the observational signatures of these superluminous Pop III Type IIn SNe \citep{wet12e}.  
There may also be hypernovae, very energetic explosions of 40 - 60 \Ms\ stars with energies 
of 10$^{52}$ erg that are intermediate to those of CC and PI SNe.  This kind of explosion, 
whose existence is inferred in part from the elemental abundances imprinted on a few of the 
most metal-poor stars found to date \citep{Iwamoto2005}, could be detected at redshifts that 
bridge those at which CC and PI SNe can be found, $15 < z < 20$.  We are also calculating 
light curves and spectra for such events.

A small number of Pop III core-collapse events may may proceed as gamma ray bursts 
\citep[GRBs; e.g.,][]{bl06,wang12}, driven either by the collapse of single rapidly-rotating 
stars \citep{suwa11,nsi12} or by binary mergers with other 20 - 50 \Ms\ stars \citep[e.g.,][]{
fw98,fwh99,zf01,pasp07}.  The prospects for mergers in particular have recently been 
strengthened by the discovery that Pop III stars sometimes form in binaries in simulations 
\citep{turk09}.  Although x-rays from these events may be detected by \textit{Swift} or its 
successors such as the \textit{Joint Astrophysics Nascent Universe Satellite} 
\citep[\textit{JANUS},][]{mesz10,Roming08,burrows10}, their afterglows \citep{wet08c} 
might also be detected in all-sky radio surveys by the \textit{Extended Very-Large Array} 
({\textit{eVLA}}), \textit{eMERLIN} and the \textit{Square Kilometer Array} (\textit{SKA}) 
\citep{ds11}.  We are now studying detection limits for Pop III GRBs in a variety of 
circumstellar environments \citep{met12a,mes13a}.  

Strong gravitational lensing by massive galaxies and clusters at $z \sim$ 0 - 1 could magnify 
flux from Pop III supernovae, improving prospects for their detection \citep{rz12}. The probability 
that flux from a Pop III SN would be boosted in an all-sky survey and its magnification depend 
on the event rate on the entire sky at a given redshift.  We have performed preliminary Markov 
Chain Monte Carlo ray-tracing calculations that suggest that the probability that a $z \sim$ 20 
event is lensed is $\sim$ 1 - 5\% for flux boosts of 2 - 5.  Much higher magnifications (10 - 300) 
are possible near the edges of massive clusters but the search volumes and probabilities of 
encountering high-$z$ SNe are much smaller.   We continue to study strong lensing of $z \sim$ 
20 events, the highest redshifts ever attempted, in order to assess its potential to discover 
primeval SNe and galaxies.  

Can later stages of the SN remnant be detected?  \citet{wet08a} found that most of the energy 
of 15 and 40 \Ms\ Pop III SNe is eventually radiated away as H and He lines as the remnant 
sweeps up and shocks pristine gas.  At lower redshifts this energy would instead be lost to fine
structure cooling by metals. In either case, the emission is too diffuse, redshifted and drawn out 
over time to be detected by any upcoming instruments.  Also, unlike PI SNe, CC SNe deposit 
little of their energy into the cosmic microwave background (CMB) by inverse Compton 
scattering at $z \sim$ 20 \citep{ky05,wet08a}, and even less at lower redshifts because the 
density of CMB photons falls with cosmological expansion.  Consequently, early populations of 
low-mass Pop III SNe will probably not impose excess power on the CMB at small scales 
\citep{oh03}.  For the same reason it will probably not be possible to directly image 
Sunyaev-Zeldovich fluctuations from individual Pop III CC SN remnants with the \textit{Atacama 
Cosmology Telescope} or the \textit{South Pole Telescope}.

Although their event rates make it unlikely that Pop III CC SNe will be detected in absorption 
at 21 cm at $z >$ 10, new calculations reveal that enough synchrotron emission from their 
remnants is redshifted into the 21 cm band above $z \sim$ 10 to be directly detected by the 
\textit{SKA} \citep{mw12}.  Somewhat more energetic hypernovae could be detected by existing 
facilities such as \textit{eVLA} and \textit{eMERLIN}.  Whether by direct detection or by their 
footprint on cosmic backgrounds, these ancient supernovae will soon open a new window on 
the $z \sim$ 10 - 15 universe.

\acknowledgments

We thank the anonymous referee, whose comments improved the quality of this paper.  DJW 
is grateful for helpful discussions with Edo Berger, Ranga Ram Chary, Daniel Kasen, Avi Loeb, 
Pete Roming and the many participants at First Stars and Galaxies:  Challenges for the Next 
Decade, held at UT Austin March 8 - 11, 2010. He also acknowledges support from the Bruce 
and Astrid McWilliams Center for Cosmology at Carnegie Mellon University. AH was supported 
by the US Department of Energy under contracts DE-FC02-01ER41176, FC02-09ER41618 
(SciDAC), and DE-FG02-87ER40328.  MS thanks Marcia Rieke for making available the 
NIRCam filter curves and was partially supported by NASA JWST grant NAG5-12458.  DEH 
acknowledges support from the National Science Foundation CAREER grant PHY-1151836.  
Work at LANL was done under the auspices of the National Nuclear Security Administration of 
the U.S. Department of Energy at Los Alamos National Laboratory under Contract No. 
DE-AC52-06NA25396. All CASTRO, RAGE and SPECTRUM calculations were performed on 
Institutional Computing (IC) and Yellow network platforms at LANL (Conejo, Lobo and Yellowrail).

\bibliographystyle{apj}
\bibliography{refs}

\begin{thebibliography}{149}
\expandafter\ifx\csname natexlab\endcsname\relax\def\natexlab#1{#1}\fi

\bibitem[{{Abel} {et~al.}(2000){Abel}, {Bryan}, \& {Norman}}]{abn00}
{Abel}, T., {Bryan}, G.~L., \& {Norman}, M.~L. 2000, \apj, 540, 39

\bibitem[{{Abel} {et~al.}(2002){Abel}, {Bryan}, \& {Norman}}]{abn02}
---. 2002, Science, 295, 93

\bibitem[{{Abel} {et~al.}(2007){Abel}, {Wise}, \& {Bryan}}]{awb07}
{Abel}, T., {Wise}, J.~H., \& {Bryan}, G.~L. 2007, \apjl, 659, L87

\bibitem[{{Agarwal} {et~al.}(2012){Agarwal}, {Khochfar}, {Johnson}, {Neistein},
  {Dalla Vecchia}, \& {Livio}}]{agarw12}
{Agarwal}, B., {Khochfar}, S., {Johnson}, J.~L., {Neistein}, E., {Dalla
  Vecchia}, C., \& {Livio}, M. 2012, \mnras, 425, 2854

\bibitem[{{Almgren} {et~al.}(2010){Almgren}, {Beckner}, {Bell}, {Day},
  {Howell}, {Joggerst}, {Lijewski}, {Nonaka}, {Singer}, \&
  {Zingale}}]{Almgren2010}
{Almgren}, A.~S., {Beckner}, V.~E., {Bell}, J.~B., {Day}, M.~S., {Howell},
  L.~H., {Joggerst}, C.~C., {Lijewski}, M.~J., {Nonaka}, A., {Singer}, M., \&
  {Zingale}, M. 2010, \apj, 715, 1221

\bibitem[{{Alvarez} {et~al.}(2006){Alvarez}, {Bromm}, \& {Shapiro}}]{abs06}
{Alvarez}, M.~A., {Bromm}, V., \& {Shapiro}, P.~R. 2006, \apj, 639, 621

\bibitem[{{Alvarez} {et~al.}(2009){Alvarez}, {Wise}, \& {Abel}}]{awa09}
{Alvarez}, M.~A., {Wise}, J.~H., \& {Abel}, T. 2009, \apjl, 701, L133

\bibitem[{{Baraffe} {et~al.}(2001){Baraffe}, {Heger}, \& {Woosley}}]{Baraffe01}
{Baraffe}, I., {Heger}, A., \& {Woosley}, S.~E. 2001, \apj, 550, 890

\bibitem[{{Beers} \& {Christlieb}(2005)}]{bc05}
{Beers}, T.~C. \& {Christlieb}, N. 2005, \araa, 43, 531

\bibitem[{{Bromm} {et~al.}(1999){Bromm}, {Coppi}, \& {Larson}}]{bcl99}
{Bromm}, V., {Coppi}, P.~S., \& {Larson}, R.~B. 1999, \apjl, 527, L5

\bibitem[{{Bromm} {et~al.}(2002){Bromm}, {Coppi}, \& {Larson}}]{bcl02}
---. 2002, \apj, 564, 23

\bibitem[{{Bromm} \& {Loeb}(2003)}]{bl03}
{Bromm}, V. \& {Loeb}, A. 2003, \apj, 596, 34

\bibitem[{{Bromm} \& {Loeb}(2006)}]{bl06}
---. 2006, \apj, 642, 382

\bibitem[{{Bromm} {et~al.}(2003){Bromm}, {Yoshida}, \& {Hernquist}}]{byh03}
{Bromm}, V., {Yoshida}, N., \& {Hernquist}, L. 2003, \apjl, 596, L135

\bibitem[{{Burrows} {et~al.}(2010){Burrows}, {Roming}, {Fox}, {Herter},
  {Falcone}, {Bil{\'e}n}, {Nousek}, \& {Kennea}}]{burrows10}
{Burrows}, D.~N., {Roming}, P.~W.~A., {Fox}, D.~B., {Herter}, T.~L., {Falcone},
  A., {Bil{\'e}n}, S., {Nousek}, J.~A., \& {Kennea}, J.~A. 2010, in Presented
  at the Society of Photo-Optical Instrumentation Engineers (SPIE) Conference,
  Vol. 7732, Society of Photo-Optical Instrumentation Engineers (SPIE)
  Conference Series

\bibitem[{{Caffau} {et~al.}(2012){Caffau}, {Bonifacio}, {Fran{\c c}ois},
  {Spite}, {Spite}, {Zaggia}, {Ludwig}, {Steffen}, {Mashonkina}, {Monaco},
  {Sbordone}, {Molaro}, {Cayrel}, {Plez}, {Hill}, {Hammer}, \&
  {Randich}}]{caffau12}
{Caffau}, E., {Bonifacio}, P., {Fran{\c c}ois}, P., {Spite}, M., {Spite}, F.,
  {Zaggia}, S., {Ludwig}, H.-G., {Steffen}, M., {Mashonkina}, L., {Monaco}, L.,
  {Sbordone}, L., {Molaro}, P., {Cayrel}, R., {Plez}, B., {Hill}, V., {Hammer},
  F., \& {Randich}, S. 2012, \aap, 542, A51

\bibitem[{{Cayrel} {et~al.}(2004){Cayrel}, {Depagne}, {Spite}, {Hill}, {Spite},
  {Fran{\c c}ois}, {Plez}, {Beers}, {Primas}, {Andersen}, {Barbuy},
  {Bonifacio}, {Molaro}, \& {Nordstr{\"o}m}}]{Cayrel2004}
{Cayrel}, R., {Depagne}, E., {Spite}, M., {Hill}, V., {Spite}, F., {Fran{\c
  c}ois}, P., {Plez}, B., {Beers}, T., {Primas}, F., {Andersen}, J., {Barbuy},
  B., {Bonifacio}, P., {Molaro}, P., \& {Nordstr{\"o}m}, B. 2004, \aap, 416,
  1117

\bibitem[{{Chatzopoulos} \& {Wheeler}(2012)}]{cw12}
{Chatzopoulos}, E. \& {Wheeler}, J.~C. 2012, \apj, 748, 42

\bibitem[{{Chen} {et~al.}(2011){Chen}, {Heger}, \& {Almgren}}]{chen11}
{Chen}, K.-J., {Heger}, A., \& {Almgren}, A.~S. 2011, Computer Physics
  Communications, 182, 254

\bibitem[{{Chiaki} {et~al.}(2013){Chiaki}, {Yoshida}, \& {Kitayama}}]{chiaki12}
{Chiaki}, G., {Yoshida}, N., \& {Kitayama}, T. 2013, \apj, 762, 50

\bibitem[{{Chieffi} \& {Limongi}(2004)}]{cl04}
{Chieffi}, A. \& {Limongi}, M. 2004, \apj, 608, 405

\bibitem[{{Clark} {et~al.}(2011){Clark}, {Glover}, {Smith}, {Greif}, {Klessen},
  \& {Bromm}}]{clark11}
{Clark}, P.~C., {Glover}, S.~C.~O., {Smith}, R.~J., {Greif}, T.~H., {Klessen},
  R.~S., \& {Bromm}, V. 2011, Science, 331, 1040

\bibitem[{{Cooke} {et~al.}(2011){Cooke}, {Pettini}, {Steidel}, {Rudie}, \&
  {Jorgenson}}]{cooke11}
{Cooke}, R., {Pettini}, M., {Steidel}, C.~C., {Rudie}, G.~C., \& {Jorgenson},
  R.~A. 2011, \mnras, 412, 1047

\bibitem[{{de Souza} {et~al.}(2011){de Souza}, {Yoshida}, \& {Ioka}}]{ds11}
{de Souza}, R.~S., {Yoshida}, N., \& {Ioka}, K. 2011, \aap, 533, A32

\bibitem[{{Dessart} {et~al.}(2013){Dessart}, {Waldman}, {Livne}, {Hillier}, \&
  {Blondin}}]{det12}
{Dessart}, L., {Waldman}, R., {Livne}, E., {Hillier}, D.~J., \& {Blondin}, S.
  2013, \mnras, 428, 3227

\bibitem[{{Djorgovski} {et~al.}(2008){Djorgovski}, {Volonteri}, {Springel},
  {Bromm}, \& {Meylan}}]{brmvol08}
{Djorgovski}, S.~G., {Volonteri}, M., {Springel}, V., {Bromm}, V., \& {Meylan},
  G. 2008, in The Eleventh Marcel Grossmann Meeting On Recent Developments in
  Theoretical and Experimental General Relativity, Gravitation and Relativistic
  Field Theories, ed. {H.~Kleinert, R.~T.~Jantzen, \& R.~Ruffini}, 340--367

\bibitem[{{Ekstr{\"o}m} {et~al.}(2008){Ekstr{\"o}m}, {Meynet}, {Chiappini},
  {Hirschi}, \& {Maeder}}]{Ekstr08}
{Ekstr{\"o}m}, S., {Meynet}, G., {Chiappini}, C., {Hirschi}, R., \& {Maeder},
  A. 2008, \aap, 489, 685

\bibitem[{{Ensman} \& {Burrows}(1992)}]{Ens92}
{Ensman}, L. \& {Burrows}, A. 1992, \apj, 393, 742

\bibitem[{{Frebel} {et~al.}(2005){Frebel}, {Aoki}, {Christlieb}, {Ando},
  {Asplund}, {Barklem}, {Beers}, {Eriksson}, {Fechner}, {Fujimoto}, {Honda},
  {Kajino}, {Minezaki}, {Nomoto}, {Norris}, {Ryan}, {Takada-Hidai},
  {Tsangarides}, \& {Yoshii}}]{fet05}
{Frebel}, A., {Aoki}, W., {Christlieb}, N., {Ando}, H., {Asplund}, M.,
  {Barklem}, P.~S., {Beers}, T.~C., {Eriksson}, K., {Fechner}, C., {Fujimoto},
  M.~Y., {Honda}, S., {Kajino}, T., {Minezaki}, T., {Nomoto}, K., {Norris},
  J.~E., {Ryan}, S.~G., {Takada-Hidai}, M., {Tsangarides}, S., \& {Yoshii}, Y.
  2005, \nat, 434, 871

\bibitem[{{Frey} {et~al.}(2013){Frey}, {Even}, {Whalen}, {Fryer}, {Hungerford},
  {Fontes}, \& {Colgan}}]{fet12}
{Frey}, L.~H., {Even}, W., {Whalen}, D.~J., {Fryer}, C.~L., {Hungerford},
  A.~L., {Fontes}, C.~J., \& {Colgan}, J. 2013, \apjs, 204, 16

\bibitem[{{Fryer} {et~al.}(2009){Fryer}, {Brown}, {Bufano}, {Dahl}, {Fontes},
  {Frey}, {Holland}, {Hungerford}, {Immler}, {Mazzali}, {Milne}, {Scannapieco},
  {Weinberg}, \& {Young}}]{fet09}
{Fryer}, C.~L., {Brown}, P.~J., {Bufano}, F., {Dahl}, J.~A., {Fontes}, C.~J.,
  {Frey}, L.~H., {Holland}, S.~T., {Hungerford}, A.~L., {Immler}, S.,
  {Mazzali}, P., {Milne}, P.~A., {Scannapieco}, E., {Weinberg}, N., \& {Young},
  P.~A. 2009, \apj, 707, 193

\bibitem[{{Fryer} {et~al.}(2007){Fryer}, {Mazzali}, {Prochaska}, {Cappellaro},
  {Panaitescu}, {Berger}, {van Putten}, {van den Heuvel}, {Young},
  {Hungerford}, {Rockefeller}, {Yoon}, {Podsiadlowski}, {Nomoto}, {Chevalier},
  {Schmidt}, \& {Kulkarni}}]{pasp07}
{Fryer}, C.~L., {Mazzali}, P.~A., {Prochaska}, J., {Cappellaro}, E.,
  {Panaitescu}, A., {Berger}, E., {van Putten}, M., {van den Heuvel}, E.~P.~J.,
  {Young}, P., {Hungerford}, A., {Rockefeller}, G., {Yoon}, S.-C.,
  {Podsiadlowski}, P., {Nomoto}, K., {Chevalier}, R., {Schmidt}, B., \&
  {Kulkarni}, S. 2007, \pasp, 119, 1211

\bibitem[{{Fryer} \& {Woosley}(1998)}]{fw98}
{Fryer}, C.~L. \& {Woosley}, S.~E. 1998, \apjl, 502, L9

\bibitem[{{Fryer} {et~al.}(1999){Fryer}, {Woosley}, \& {Hartmann}}]{fwh99}
{Fryer}, C.~L., {Woosley}, S.~E., \& {Hartmann}, D.~H. 1999, \apj, 526, 152

\bibitem[{{Gal-Yam} {et~al.}(2009){Gal-Yam}, {Mazzali}, {Ofek}, {Nugent},
  {Kulkarni}, {Kasliwal}, {Quimby}, {Filippenko}, {Cenko}, {Chornock},
  {Waldman}, {Kasen}, {Sullivan}, {Beshore}, {Drake}, {Thomas}, {Bloom},
  {Poznanski}, {Miller}, {Foley}, {Silverman}, {Arcavi}, {Ellis}, \&
  {Deng}}]{gy09}
{Gal-Yam}, A., {Mazzali}, P., {Ofek}, E.~O., {Nugent}, P.~E., {Kulkarni},
  S.~R., {Kasliwal}, M.~M., {Quimby}, R.~M., {Filippenko}, A.~V., {Cenko},
  S.~B., {Chornock}, R., {Waldman}, R., {Kasen}, D., {Sullivan}, M., {Beshore},
  E.~C., {Drake}, A.~J., {Thomas}, R.~C., {Bloom}, J.~S., {Poznanski}, D.,
  {Miller}, A.~A., {Foley}, R.~J., {Silverman}, J.~M., {Arcavi}, I., {Ellis},
  R.~S., \& {Deng}, J. 2009, \nat, 462, 624

\bibitem[{{Gardner} {et~al.}(2006){Gardner}, {Mather}, {Clampin}, {Doyon},
  {Greenhouse}, {Hammel}, {Hutchings}, {Jakobsen}, {Lilly}, {Long}, {Lunine},
  {McCaughrean}, {Mountain}, {Nella}, {Rieke}, {Rieke}, {Rix}, {Smith},
  {Sonneborn}, {Stiavelli}, {Stockman}, {Windhorst}, \& {Wright}}]{jwst06}
{Gardner}, J.~P., {Mather}, J.~C., {Clampin}, M., {Doyon}, R., {Greenhouse},
  M.~A., {Hammel}, H.~B., {Hutchings}, J.~B., {Jakobsen}, P., {Lilly}, S.~J.,
  {Long}, K.~S., {Lunine}, J.~I., {McCaughrean}, M.~J., {Mountain}, M.,
  {Nella}, J., {Rieke}, G.~H., {Rieke}, M.~J., {Rix}, H.-W., {Smith}, E.~P.,
  {Sonneborn}, G., {Stiavelli}, M., {Stockman}, H.~S., {Windhorst}, R.~A., \&
  {Wright}, G.~S. 2006, \ssr, 123, 485

\bibitem[{{Gittings} {et~al.}(2008){Gittings}, {Weaver}, {Clover}, {Betlach},
  {Byrne}, {Coker}, {Dendy}, {Hueckstaedt}, {New}, {Oakes}, {Ranta}, \&
  {Stefan}}]{rage}
{Gittings}, M., {Weaver}, R., {Clover}, M., {Betlach}, T., {Byrne}, N.,
  {Coker}, R., {Dendy}, E., {Hueckstaedt}, R., {New}, K., {Oakes}, W.~R.,
  {Ranta}, D., \& {Stefan}, R. 2008, Computational Science and Discovery, 1,
  015005

\bibitem[{{Glover}(2012)}]{glov12}
{Glover}, S.~C.~O. 2012, arXiv:1209.2509

\bibitem[{{Greif} {et~al.}(2012){Greif}, {Bromm}, {Clark}, {Glover}, {Smith},
  {Klessen}, {Yoshida}, \& {Springel}}]{get12}
{Greif}, T.~H., {Bromm}, V., {Clark}, P.~C., {Glover}, S.~C.~O., {Smith},
  R.~J., {Klessen}, R.~S., {Yoshida}, N., \& {Springel}, V. 2012, \mnras, 424,
  399

\bibitem[{{Greif} {et~al.}(2010){Greif}, {Glover}, {Bromm}, \&
  {Klessen}}]{get10}
{Greif}, T.~H., {Glover}, S.~C.~O., {Bromm}, V., \& {Klessen}, R.~S. 2010,
  \apj, 716, 510

\bibitem[{{Greif} {et~al.}(2008){Greif}, {Johnson}, {Klessen}, \&
  {Bromm}}]{get08}
{Greif}, T.~H., {Johnson}, J.~L., {Klessen}, R.~S., \& {Bromm}, V. 2008,
  \mnras, 387, 1021

\bibitem[{{Greif} {et~al.}(2011){Greif}, {Springel}, {White}, {Glover},
  {Clark}, {Smith}, {Klessen}, \& {Bromm}}]{get11}
{Greif}, T.~H., {Springel}, V., {White}, S.~D.~M., {Glover}, S.~C.~O., {Clark},
  P.~C., {Smith}, R.~J., {Klessen}, R.~S., \& {Bromm}, V. 2011, \apj, 737, 75

\bibitem[{{Heger} \& {Woosley}(2002)}]{hw02}
{Heger}, A. \& {Woosley}, S.~E. 2002, \apj, 567, 532

\bibitem[{{Hosokawa} {et~al.}(2011){Hosokawa}, {Omukai}, {Yoshida}, \&
  {Yorke}}]{hos11}
{Hosokawa}, T., {Omukai}, K., {Yoshida}, N., \& {Yorke}, H.~W. 2011, Science,
  334, 1250

\bibitem[{{Hosokawa} {et~al.}(2012){Hosokawa}, {Yoshida}, {Omukai}, \&
  {Yorke}}]{hos12}
{Hosokawa}, T., {Yoshida}, N., {Omukai}, K., \& {Yorke}, H.~W. 2012, \apjl,
  760, L37

\bibitem[{{Hummel} {et~al.}(2012){Hummel}, {Pawlik}, {Milosavljevi{\'c}}, \&
  {Bromm}}]{hum12}
{Hummel}, J.~A., {Pawlik}, A.~H., {Milosavljevi{\'c}}, M., \& {Bromm}, V. 2012,
  \apj, 755, 72

\bibitem[{{Iwamoto} {et~al.}(2005){Iwamoto}, {Umeda}, {Tominaga}, {Nomoto}, \&
  {Maeda}}]{Iwamoto2005}
{Iwamoto}, N., {Umeda}, H., {Tominaga}, N., {Nomoto}, K., \& {Maeda}, K. 2005,
  Science, 309, 451

\bibitem[{{Jeon} {et~al.}(2012){Jeon}, {Pawlik}, {Greif}, {Glover}, {Bromm},
  {Milosavljevi{\'c}}, \& {Klessen}}]{jeon11}
{Jeon}, M., {Pawlik}, A.~H., {Greif}, T.~H., {Glover}, S.~C.~O., {Bromm}, V.,
  {Milosavljevi{\'c}}, M., \& {Klessen}, R.~S. 2012, \apj, 754, 34

\bibitem[{{Joggerst} {et~al.}(2010){Joggerst}, {Almgren}, {Bell}, {Heger},
  {Whalen}, \& {Woosley}}]{jet09b}
{Joggerst}, C.~C., {Almgren}, A., {Bell}, J., {Heger}, A., {Whalen}, D., \&
  {Woosley}, S.~E. 2010, \apj, 709, 11

\bibitem[{{Joggerst} \& {Whalen}(2011)}]{jw11}
{Joggerst}, C.~C. \& {Whalen}, D.~J. 2011, \apj, 728, 129

\bibitem[{{Johnson} \& {Bromm}(2007)}]{jb07b}
{Johnson}, J.~L. \& {Bromm}, V. 2007, \mnras, 374, 1557

\bibitem[{{Johnson} {et~al.}(2013){Johnson}, {Dalla}, \& {Khochfar}}]{jdk12}
{Johnson}, J.~L., {Dalla}, V.~C., \& {Khochfar}, S. 2013, \mnras, 428, 1857

\bibitem[{{Johnson} {et~al.}(2008){Johnson}, {Greif}, \& {Bromm}}]{jgb08}
{Johnson}, J.~L., {Greif}, T.~H., \& {Bromm}, V. 2008, \mnras, 388, 26

\bibitem[{{Johnson} {et~al.}(2009){Johnson}, {Greif}, {Bromm}, {Klessen}, \&
  {Ippolito}}]{jlj09}
{Johnson}, J.~L., {Greif}, T.~H., {Bromm}, V., {Klessen}, R.~S., \& {Ippolito},
  J. 2009, \mnras, 399, 37

\bibitem[{{Johnson} {et~al.}(2012{\natexlab{a}}){Johnson}, {Whalen}, {Fryer},
  \& {Li}}]{jlj12a}
{Johnson}, J.~L., {Whalen}, D.~J., {Fryer}, C.~L., \& {Li}, H.
  2012{\natexlab{a}}, \apj, 750, 66

\bibitem[{{Johnson} {et~al.}(2012{\natexlab{b}}){Johnson}, {Whalen}, {Li}, \&
  {Holz}}]{jet13}
{Johnson}, J.~L., {Whalen}, D.~J., {Li}, H., \& {Holz}, D.~E.
  2012{\natexlab{b}}, arXiv:1211.0548

\bibitem[{{Karlsson} {et~al.}(2008){Karlsson}, {Johnson}, \& {Bromm}}]{karl08}
{Karlsson}, T., {Johnson}, J.~L., \& {Bromm}, V. 2008, \apj, 679, 6

\bibitem[{{Kasen} {et~al.}(2011){Kasen}, {Woosley}, \& {Heger}}]{kasen11}
{Kasen}, D., {Woosley}, S.~E., \& {Heger}, A. 2011, \apj, 734, 102

\bibitem[{{Kitayama} \& {Yoshida}(2005)}]{ky05}
{Kitayama}, T. \& {Yoshida}, N. 2005, \apj, 630, 675

\bibitem[{{Kitayama} {et~al.}(2004){Kitayama}, {Yoshida}, {Susa}, \&
  {Umemura}}]{ket04}
{Kitayama}, T., {Yoshida}, N., {Susa}, H., \& {Umemura}, M. 2004, \apj, 613,
  631

\bibitem[{{Klein} \& {Chevalier}(1978)}]{kl78}
{Klein}, R.~I. \& {Chevalier}, R.~A. 1978, \apjl, 223, L109

\bibitem[{{Krti{\v c}ka} \& {Kub{\'a}t}(2006)}]{kk06}
{Krti{\v c}ka}, J. \& {Kub{\'a}t}, J. 2006, \aap, 446, 1039

\bibitem[{{Kudritzki}(2000)}]{Kudritzki00}
{Kudritzki}, R. 2000, in The First Stars, ed. {A.~Weiss, T.~G.~Abel, \&
  V.~Hill}, 127--+

\bibitem[{{Lai} {et~al.}(2008){Lai}, {Bolte}, {Johnson}, {Lucatello}, {Heger},
  \& {Woosley}}]{Lai2008}
{Lai}, D.~K., {Bolte}, M., {Johnson}, J.~A., {Lucatello}, S., {Heger}, A., \&
  {Woosley}, S.~E. 2008, \apj, 681, 1524

\bibitem[{{Latif} {et~al.}(2013){Latif}, {Schleicher}, {Schmidt}, \&
  {Niemeyer}}]{schl13}
{Latif}, M.~A., {Schleicher}, D.~R.~G., {Schmidt}, W., \& {Niemeyer}, J. 2013,
  \mnras, 551

\bibitem[{{Li}(2011)}]{li11}
{Li}, Y. 2011, arXiv:1109.3442

\bibitem[{{Lippai} {et~al.}(2009){Lippai}, {Frei}, \& {Haiman}}]{lfh09}
{Lippai}, Z., {Frei}, Z., \& {Haiman}, Z. 2009, \apj, 701, 360

\bibitem[{{Mackey} {et~al.}(2003){Mackey}, {Bromm}, \& {Hernquist}}]{mbh03}
{Mackey}, J., {Bromm}, V., \& {Hernquist}, L. 2003, \apj, 586, 1

\bibitem[{{Madau}(1995)}]{madau95}
{Madau}, P. 1995, \apj, 441, 18

\bibitem[{{Magee} {et~al.}(1995){Magee}, {Abdallah}, {Clark}, {Cohen},
  {Collins}, {Csanak}, {Fontes}, {Gauger}, {Keady}, {Kilcrease}, \&
  {Merts}}]{oplib}
{Magee}, N.~H., {Abdallah}, Jr., J., {Clark}, R.~E.~H., {Cohen}, J.~S.,
  {Collins}, L.~A., {Csanak}, G., {Fontes}, C.~J., {Gauger}, A., {Keady},
  J.~J., {Kilcrease}, D.~P., \& {Merts}, A.~L. 1995, in Astronomical Society of
  the Pacific Conference Series, Vol.~78, Astrophysical Applications of
  Powerful New Databases, ed. {S.~J.~Adelman \& W.~L.~Wiese}, 51

\bibitem[{{Maio} {et~al.}(2011){Maio}, {Khochfar}, {Johnson}, \&
  {Ciardi}}]{maio11}
{Maio}, U., {Khochfar}, S., {Johnson}, J.~L., \& {Ciardi}, B. 2011, \mnras,
  414, 1145

\bibitem[{{McKee} \& {Tan}(2008)}]{tm08}
{McKee}, C.~F. \& {Tan}, J.~C. 2008, \apj, 681, 771

\bibitem[{{Meiksin} \& {Whalen}(2012)}]{mw12}
{Meiksin}, A. \& {Whalen}, D.~J. 2012, arXiv:1209.1915

\bibitem[{{Mesler} {et~al.}(2012){Mesler}, {Whalen}, {Lloyd-Ronning}, {Fryer},
  \& {Pihlstr{\"o}m}}]{met12a}
{Mesler}, R.~A., {Whalen}, D.~J., {Lloyd-Ronning}, N.~M., {Fryer}, C.~L., \&
  {Pihlstr{\"o}m}, Y.~M. 2012, \apj, 757, 117

\bibitem[{{Mesler} {et~al.}(2013){Mesler}, {Whalen}, {Lloyd-Ronning}, {Fryer},
  \& {Pihlstr{\"o}m}}]{mes13a}
---. 2013, \apj, in prep

\bibitem[{{M{\'e}sz{\'a}ros} \& {Rees}(2010)}]{mesz10}
{M{\'e}sz{\'a}ros}, P. \& {Rees}, M.~J. 2010, \apj, 715, 967

\bibitem[{{Milosavljevi{\'c}} {et~al.}(2009){Milosavljevi{\'c}}, {Bromm},
  {Couch}, \& {Oh}}]{milos09}
{Milosavljevi{\'c}}, M., {Bromm}, V., {Couch}, S.~M., \& {Oh}, S.~P. 2009,
  \apj, 698, 766

\bibitem[{{Moriya} {et~al.}(2010){Moriya}, {Yoshida}, {Tominaga}, {Blinnikov},
  {Maeda}, {Tanaka}, \& {Nomoto}}]{moriya10}
{Moriya}, T., {Yoshida}, N., {Tominaga}, N., {Blinnikov}, S.~I., {Maeda}, K.,
  {Tanaka}, M., \& {Nomoto}, K. 2010, in American Institute of Physics
  Conference Series, Vol. 1294, American Institute of Physics Conference
  Series, ed. {D.~J.~Whalen, V.~Bromm, \& N.~Yoshida}, 268--269

\bibitem[{{Moriya} {et~al.}(2013){Moriya}, {Blinnikov}, {Tominaga}, {Yoshida},
  {Tanaka}, {Maeda}, \& {Nomoto}}]{moriya12}
{Moriya}, T.~J., {Blinnikov}, S.~I., {Tominaga}, N., {Yoshida}, N., {Tanaka},
  M., {Maeda}, K., \& {Nomoto}, K. 2013, \mnras, 428, 1020

\bibitem[{{Nagakura} {et~al.}(2012){Nagakura}, {Suwa}, \& {Ioka}}]{nsi12}
{Nagakura}, H., {Suwa}, Y., \& {Ioka}, K. 2012, \apj, 754, 85

\bibitem[{{Nakamura} \& {Umemura}(2001)}]{nu01}
{Nakamura}, F. \& {Umemura}, M. 2001, \apj, 548, 19

\bibitem[{{Oh} {et~al.}(2003){Oh}, {Cooray}, \& {Kamionkowski}}]{oh03}
{Oh}, S.~P., {Cooray}, A., \& {Kamionkowski}, M. 2003, \mnras, 342, L20

\bibitem[{{Ohyama}(1963)}]{ohy63}
{Ohyama}, N. 1963, Progress of Theoretical Physics, 30, 170

\bibitem[{{Omukai} \& {Inutsuka}(2002)}]{oi02}
{Omukai}, K. \& {Inutsuka}, S.-i. 2002, \mnras, 332, 59

\bibitem[{{Omukai} \& {Palla}(2001)}]{op01}
{Omukai}, K. \& {Palla}, F. 2001, \apjl, 561, L55

\bibitem[{{Omukai} \& {Palla}(2003)}]{op03}
---. 2003, \apj, 589, 677

\bibitem[{{O'Shea} {et~al.}(2005){O'Shea}, {Abel}, {Whalen}, \&
  {Norman}}]{oet05}
{O'Shea}, B.~W., {Abel}, T., {Whalen}, D., \& {Norman}, M.~L. 2005, \apjl, 628,
  L5

\bibitem[{{O'Shea} \& {Norman}(2007)}]{on07}
{O'Shea}, B.~W. \& {Norman}, M.~L. 2007, \apj, 654, 66

\bibitem[{{O'Shea} \& {Norman}(2008)}]{on08}
---. 2008, \apj, 673, 14

\bibitem[{{Pan} {et~al.}(2012{\natexlab{a}}){Pan}, {Kasen}, \& {Loeb}}]{pan12a}
{Pan}, T., {Kasen}, D., \& {Loeb}, A. 2012{\natexlab{a}}, \mnras, 422, 2701

\bibitem[{{Pan} {et~al.}(2012{\natexlab{b}}){Pan}, {Loeb}, \& {Kasen}}]{pan12b}
{Pan}, T., {Loeb}, A., \& {Kasen}, D. 2012{\natexlab{b}}, \mnras, 423, 2203

\bibitem[{{Park} \& {Ricotti}(2011)}]{pm11}
{Park}, K. \& {Ricotti}, M. 2011, \apj, 739, 2

\bibitem[{{Park} \& {Ricotti}(2012{\natexlab{a}})}]{pm12}
---. 2012{\natexlab{a}}, \apj, 747, 9

\bibitem[{{Park} \& {Ricotti}(2012{\natexlab{b}})}]{pm13}
---. 2012{\natexlab{b}}, arXiv:1211.0542

\bibitem[{{Pawlik} {et~al.}(2011){Pawlik}, {Milosavljevi{\'c}}, \&
  {Bromm}}]{pmb11}
{Pawlik}, A.~H., {Milosavljevi{\'c}}, M., \& {Bromm}, V. 2011, \apj, 731, 54

\bibitem[{{Pawlik} {et~al.}(2012){Pawlik}, {Milosavljevic}, \& {Bromm}}]{pmb12}
{Pawlik}, A.~H., {Milosavljevic}, M., \& {Bromm}, V. 2012, arXiv:1208.3698

\bibitem[{{Ren} {et~al.}(2012){Ren}, {Christlieb}, \& {Zhao}}]{ren12}
{Ren}, J., {Christlieb}, N., \& {Zhao}, G. 2012, Research in Astronomy and
  Astrophysics, 12, 1637

\bibitem[{{Ritter} {et~al.}(2012){Ritter}, {Safranek-Shrader}, {Gnat},
  {Milosavljevi{\'c}}, \& {Bromm}}]{ritt12}
{Ritter}, J.~S., {Safranek-Shrader}, C., {Gnat}, O., {Milosavljevi{\'c}}, M.,
  \& {Bromm}, V. 2012, \apj, 761, 56

\bibitem[{{Roming}(2008)}]{Roming08}
{Roming}, P. 2008, in COSPAR, Plenary Meeting, Vol.~37, 37th COSPAR Scientific
  Assembly, 2645--+

\bibitem[{{Rydberg} {et~al.}(2013){Rydberg}, {Zackrisson}, {Lundqvist}, \&
  {Scott}}]{rz12}
{Rydberg}, C.-E., {Zackrisson}, E., {Lundqvist}, P., \& {Scott}, P. 2013,
  \mnras, 574

\bibitem[{{Scannapieco} {et~al.}(2005){Scannapieco}, {Madau}, {Woosley},
  {Heger}, \& {Ferrara}}]{sc05}
{Scannapieco}, E., {Madau}, P., {Woosley}, S., {Heger}, A., \& {Ferrara}, A.
  2005, \apj, 633, 1031

\bibitem[{{Schober} {et~al.}(2012){Schober}, {Schleicher}, {Federrath},
  {Glover}, {Klessen}, \& {Banerjee}}]{schob12}
{Schober}, J., {Schleicher}, D., {Federrath}, C., {Glover}, S., {Klessen},
  R.~S., \& {Banerjee}, R. 2012, \apj, 754, 99

\bibitem[{{Smith} \& {Sigurdsson}(2007)}]{ss07}
{Smith}, B.~D. \& {Sigurdsson}, S. 2007, \apjl, 661, L5

\bibitem[{{Smith} {et~al.}(2009){Smith}, {Turk}, {Sigurdsson}, {O'Shea}, \&
  {Norman}}]{bsmith09}
{Smith}, B.~D., {Turk}, M.~J., {Sigurdsson}, S., {O'Shea}, B.~W., \& {Norman},
  M.~L. 2009, \apj, 691, 441

\bibitem[{{Smith} {et~al.}(2007){Smith}, {Li}, {Foley}, {Wheeler}, {Pooley},
  {Chornock}, {Filippenko}, {Silverman}, {Quimby}, {Bloom}, \&
  {Hansen}}]{nsmith07b}
{Smith}, N., {Li}, W., {Foley}, R.~J., {Wheeler}, J.~C., {Pooley}, D.,
  {Chornock}, R., {Filippenko}, A.~V., {Silverman}, J.~M., {Quimby}, R.,
  {Bloom}, J.~S., \& {Hansen}, C. 2007, \apj, 666, 1116

\bibitem[{{Smith} \& {McCray}(2007)}]{nsmith07a}
{Smith}, N. \& {McCray}, R. 2007, \apjl, 671, L17

\bibitem[{{Smith} {et~al.}(2011){Smith}, {Glover}, {Clark}, {Greif}, \&
  {Klessen}}]{sm11}
{Smith}, R.~J., {Glover}, S.~C.~O., {Clark}, P.~C., {Greif}, T., \& {Klessen},
  R.~S. 2011, \mnras, 414, 3633

\bibitem[{{Stacy} {et~al.}(2010){Stacy}, {Greif}, \& {Bromm}}]{stacy10}
{Stacy}, A., {Greif}, T.~H., \& {Bromm}, V. 2010, \mnras, 403, 45

\bibitem[{{Stacy} {et~al.}(2012){Stacy}, {Greif}, \& {Bromm}}]{stacy12}
---. 2012, \mnras, 422, 290

\bibitem[{{Su} {et~al.}(2011){Su}, {Stiavelli}, {Oesch}, {Trenti}, {Bergeron},
  {Bradley}, {Carollo}, {Dahlen}, {Ferguson}, {Giavalisco}, {Koekemoer},
  {Lilly}, {Lucas}, {Mobasher}, {Panagia}, \& {Pavlovsky}}]{su11}
{Su}, J., {Stiavelli}, M., {Oesch}, P., {Trenti}, M., {Bergeron}, E.,
  {Bradley}, L., {Carollo}, M., {Dahlen}, T., {Ferguson}, H.~C., {Giavalisco},
  M., {Koekemoer}, A., {Lilly}, S., {Lucas}, R.~A., {Mobasher}, B., {Panagia},
  N., \& {Pavlovsky}, C. 2011, \apj, 738, 123

\bibitem[{{Suwa} \& {Ioka}(2011)}]{suwa11}
{Suwa}, Y. \& {Ioka}, K. 2011, \apj, 726, 107

\bibitem[{{Tan} \& {McKee}(2004)}]{tm04}
{Tan}, J.~C. \& {McKee}, C.~F. 2004, \apj, 603, 383

\bibitem[{{Tanaka} {et~al.}(2012){Tanaka}, {Moriya}, {Yoshida}, \&
  {Nomoto}}]{tet12}
{Tanaka}, M., {Moriya}, T.~J., {Yoshida}, N., \& {Nomoto}, K. 2012, \mnras,
  422, 2675

\bibitem[{{Tanaka} \& {Haiman}(2009)}]{th09}
{Tanaka}, T. \& {Haiman}, Z. 2009, \apj, 696, 1798

\bibitem[{{Tolstov} {et~al.}(2013){Tolstov}, {Blinnikov}, \&
  {Nadyozhin}}]{tbn13}
{Tolstov}, A.~G., {Blinnikov}, S.~I., \& {Nadyozhin}, D.~K. 2013, \mnras, 429,
  3181

\bibitem[{{Tominaga} {et~al.}(2011){Tominaga}, {Morokuma}, {Blinnikov},
  {Baklanov}, {Sorokina}, \& {Nomoto}}]{tomin11}
{Tominaga}, N., {Morokuma}, T., {Blinnikov}, S.~I., {Baklanov}, P., {Sorokina},
  E.~I., \& {Nomoto}, K. 2011, \apjs, 193, 20

\bibitem[{{Tornatore} {et~al.}(2007){Tornatore}, {Ferrara}, \&
  {Schneider}}]{tfs07}
{Tornatore}, L., {Ferrara}, A., \& {Schneider}, R. 2007, \mnras, 382, 945

\bibitem[{{Trenti} {et~al.}(2009){Trenti}, {Stiavelli}, \& {Michael
  Shull}}]{tss09}
{Trenti}, M., {Stiavelli}, M., \& {Michael Shull}, J. 2009, \apj, 700, 1672

\bibitem[{{Turk} {et~al.}(2009){Turk}, {Abel}, \& {O'Shea}}]{turk09}
{Turk}, M.~J., {Abel}, T., \& {O'Shea}, B. 2009, Science, 325, 601

\bibitem[{{van Marle} {et~al.}(2010){van Marle}, {Smith}, {Owocki}, \& {van
  Veelen}}]{vmarle10}
{van Marle}, A.~J., {Smith}, N., {Owocki}, S.~P., \& {van Veelen}, B. 2010,
  \mnras, 407, 2305

\bibitem[{{Vasiliev} {et~al.}(2012){Vasiliev}, {Vorobyov}, {Matvienko},
  {Razoumov}, \& {Shchekinov}}]{vas12}
{Vasiliev}, E.~O., {Vorobyov}, E.~I., {Matvienko}, E.~E., {Razoumov}, A.~O., \&
  {Shchekinov}, Y.~A. 2012, Astronomy Reports, 56, 895

\bibitem[{{Vink} {et~al.}(2001){Vink}, {de Koter}, \& {Lamers}}]{Vink01}
{Vink}, J.~S., {de Koter}, A., \& {Lamers}, H.~J.~G.~L.~M. 2001, \aap, 369, 574

\bibitem[{{Wang} {et~al.}(2012){Wang}, {Bromm}, {Greif}, {Stacy}, {Dai},
  {Loeb}, \& {Cheng}}]{wang12}
{Wang}, F.~Y., {Bromm}, V., {Greif}, T.~H., {Stacy}, A., {Dai}, Z.~G., {Loeb},
  A., \& {Cheng}, K.~S. 2012, \apj, 760, 27

\bibitem[{{Weaver} {et~al.}(1978){Weaver}, {Zimmerman}, \&
  {Woosley}}]{Weaver1978}
{Weaver}, T.~A., {Zimmerman}, G.~B., \& {Woosley}, S.~E. 1978, \apj, 225, 1021

\bibitem[{{Weinmann} \& {Lilly}(2005)}]{wl05}
{Weinmann}, S.~M. \& {Lilly}, S.~J. 2005, \apj, 624, 526

\bibitem[{{Whalen} {et~al.}(2004){Whalen}, {Abel}, \& {Norman}}]{wan04}
{Whalen}, D., {Abel}, T., \& {Norman}, M.~L. 2004, \apj, 610, 14

\bibitem[{{Whalen} {et~al.}(2010){Whalen}, {Hueckstaedt}, \&
  {McConkie}}]{wet10}
{Whalen}, D., {Hueckstaedt}, R.~M., \& {McConkie}, T.~O. 2010, \apj, 712, 101

\bibitem[{{Whalen} \& {Norman}(2006)}]{wn06}
{Whalen}, D. \& {Norman}, M.~L. 2006, \apjs, 162, 281

\bibitem[{{Whalen} \& {Norman}(2008{\natexlab{a}})}]{wn08b}
---. 2008{\natexlab{a}}, \apj, 673, 664

\bibitem[{{Whalen} {et~al.}(2008{\natexlab{a}}){Whalen}, {O'Shea}, {Smidt}, \&
  {Norman}}]{wet08b}
{Whalen}, D., {O'Shea}, B.~W., {Smidt}, J., \& {Norman}, M.~L.
  2008{\natexlab{a}}, \apj, 679, 925

\bibitem[{{Whalen} {et~al.}(2008{\natexlab{b}}){Whalen}, {Prochaska}, {Heger},
  \& {Tumlinson}}]{wet08c}
{Whalen}, D., {Prochaska}, J.~X., {Heger}, A., \& {Tumlinson}, J.
  2008{\natexlab{b}}, \apj, 682, 1114

\bibitem[{{Whalen} {et~al.}(2008{\natexlab{c}}){Whalen}, {van Veelen},
  {O'Shea}, \& {Norman}}]{wet08a}
{Whalen}, D., {van Veelen}, B., {O'Shea}, B.~W., \& {Norman}, M.~L.
  2008{\natexlab{c}}, \apj, 682, 49

\bibitem[{{Whalen}(2012)}]{dw12}
{Whalen}, D.~J. 2012, arXiv:1209.4688

\bibitem[{{Whalen} {et~al.}(2012{\natexlab{a}}){Whalen}, {Even}, {Frey},
  {Johnson}, {Lovekin}, {Fryer}, {Stiavelli}, {Holz}, {Heger}, {Woosley}, \&
  {Hungerford}}]{wet12b}
{Whalen}, D.~J., {Even}, W., {Frey}, L.~H., {Johnson}, J.~L., {Lovekin}, C.~C.,
  {Fryer}, C.~L., {Stiavelli}, M., {Holz}, D.~E., {Heger}, A., {Woosley},
  S.~E., \& {Hungerford}, A.~L. 2012{\natexlab{a}}, arXiv:1211.4979

\bibitem[{{Whalen} {et~al.}(2013{\natexlab{a}}){Whalen}, {Even}, {Lovekin},
  {Fryer}, {Stiavelli}, {Roming}, {Cooke}, {Pritchard}, {Holz}, \&
  {Knight}}]{wet12e}
{Whalen}, D.~J., {Even}, W., {Lovekin}, C.~C., {Fryer}, C.~L., {Stiavelli}, M.,
  {Roming}, P.~W.~A., {Cooke}, J., {Pritchard}, T.~A., {Holz}, D.~E., \&
  {Knight}, C. 2013{\natexlab{a}}, arXiv:1302.0436

\bibitem[{{Whalen} \& {Fryer}(2012)}]{wf12}
{Whalen}, D.~J. \& {Fryer}, C.~L. 2012, \apjl, 756, L19

\bibitem[{{Whalen} {et~al.}(2013{\natexlab{b}}){Whalen}, {Fryer}, {Holz},
  {Heger}, {Woosley}, {Stiavelli}, {Even}, \& {Frey}}]{wet12a}
{Whalen}, D.~J., {Fryer}, C.~L., {Holz}, D.~E., {Heger}, A., {Woosley}, S.~E.,
  {Stiavelli}, M., {Even}, W., \& {Frey}, L.~H. 2013{\natexlab{b}}, \apjl, 762,
  L6

\bibitem[{{Whalen} {et~al.}(2012{\natexlab{b}}){Whalen}, {Heger}, {Chen},
  {Even}, {Fryer}, {Stiavelli}, {Xu}, \& {Joggerst}}]{wet12d}
{Whalen}, D.~J., {Heger}, A., {Chen}, K.-J., {Even}, W., {Fryer}, C.~L.,
  {Stiavelli}, M., {Xu}, H., \& {Joggerst}, C.~C. 2012{\natexlab{b}},
  arXiv:1211.1815

\bibitem[{{Whalen} \& {Norman}(2008{\natexlab{b}})}]{wn08a}
{Whalen}, D.~J. \& {Norman}, M.~L. 2008{\natexlab{b}}, \apj, 672, 287

\bibitem[{{Wise} \& {Abel}(2005)}]{wa05}
{Wise}, J.~H. \& {Abel}, T. 2005, \apj, 629, 615

\bibitem[{{Wise} \& {Abel}(2007)}]{wa07}
---. 2007, \apj, 671, 1559

\bibitem[{{Wise} \& {Abel}(2008)}]{wa08a}
---. 2008, \apj, 684, 1

\bibitem[{{Wise} {et~al.}(2012){Wise}, {Turk}, {Norman}, \& {Abel}}]{wise12}
{Wise}, J.~H., {Turk}, M.~J., {Norman}, M.~L., \& {Abel}, T. 2012, \apj, 745,
  50

\bibitem[{{Woosley} \& {Heger}(2007)}]{wh07}
{Woosley}, S.~E. \& {Heger}, A. 2007, \physrep, 442, 269

\bibitem[{{Woosley} {et~al.}(2002){Woosley}, {Heger}, \&
  {Weaver}}]{Woosley2002}
{Woosley}, S.~E., {Heger}, A., \& {Weaver}, T.~A. 2002, Reviews of Modern
  Physics, 74, 1015

\bibitem[{{Yoshida} {et~al.}(2008){Yoshida}, {Omukai}, \& {Hernquist}}]{y08}
{Yoshida}, N., {Omukai}, K., \& {Hernquist}, L. 2008, Science, 321, 669

\bibitem[{{Young} {et~al.}(2010){Young}, {Smartt}, {Valenti}, {Pastorello},
  {Benetti}, {Benn}, {Bersier}, {Botticella}, {Corradi}, {Harutyunyan},
  {Hrudkova}, {Hunter}, {Mattila}, {de Mooij}, {Navasardyan}, {Snellen},
  {Tanvir}, \& {Zampieri}}]{yn10}
{Young}, D.~R., {Smartt}, S.~J., {Valenti}, S., {Pastorello}, A., {Benetti},
  S., {Benn}, C.~R., {Bersier}, D., {Botticella}, M.~T., {Corradi}, R.~L.~M.,
  {Harutyunyan}, A.~H., {Hrudkova}, M., {Hunter}, I., {Mattila}, S., {de
  Mooij}, E.~J.~W., {Navasardyan}, H., {Snellen}, I.~A.~G., {Tanvir}, N.~R., \&
  {Zampieri}, L. 2010, \aap, 512, A70+

\bibitem[{{Zhang} \& {Fryer}(2001)}]{zf01}
{Zhang}, W. \& {Fryer}, C.~L. 2001, \apj, 550, 357

\bibitem[{{Zhang} {et~al.}(2008){Zhang}, {Woosley}, \& {Heger}}]{zwh08}
{Zhang}, W., {Woosley}, S.~E., \& {Heger}, A. 2008, \apj, 679, 639

\end{thebibliography}

\end{document}